\documentclass[a4paper,11pt]{article}
\usepackage{jheppub}

\usepackage{bm}
\usepackage{enumerate}
\usepackage{graphicx}
\usepackage{subfigure}
\usepackage{braket}

\usepackage{array}
\usepackage{setspace}

\usepackage{hyperref}

\usepackage{amsmath,amssymb,amsfonts,amsthm,latexsym,stmaryrd}

\newcommand{\be}{\begin{equation}}
\newcommand{\ee}{\end{equation}}
\newcommand{\beq}{\begin{eqnarray}}
\newcommand{\eeq}{\end{eqnarray}}
\newcommand{\bea}{\begin{eqnarray}}
\newcommand{\eea}{\end{eqnarray}}
\newcommand{\beqn}{\begin{eqnarray}}
\newcommand{\eeqn}{\end{eqnarray}}

\def\n{\bm{n}}
\def\s{\bm{s}}
\DeclareMathOperator{\Ker}{Ker}

\def\si{{\sigma}}

\def\rd{\mathrm{d}}
\def\pa{\partial}

\newcommand{\hateq}{\mathrel{\hat{=}}}

\newcommand{\SO}{\text{SO}}
\newcommand{\R}{\mathbb{R}}

\newcommand{\Tr}{\mbox{Tr}}

\newcommand{\slr}{\text{SL}(2,\mathbb{R})}

\begin{document}

\title{Local subsystems in gauge theory and gravity}

\author[a]{William Donnelly}
\emailAdd{donnelly@physics.ucsb.edu}
\affiliation[a]{Department of Physics, University of California, Santa Barbara \\
Santa Barbara, California 93106, USA}

\author[b]{Laurent Freidel}
\emailAdd{lfreidel@perimeterinstitute.ca}
\affiliation[b]{Perimeter Institute for Theoretical Physics\\
31 Caroline St. N, N2L 2Y5, Waterloo ON, Canada}

\date{\today}

\abstract{
We consider the problem of defining localized subsystems in gauge theory and gravity.
Such systems are associated to spacelike hypersurfaces with boundaries and provide the natural setting for studying entanglement entropy of localized subsystems.
We present a general formalism to associate a gauge-invariant classical phase space to a spatial slice with boundary by introducing new degrees of freedom on the boundary.
In Yang-Mills theory the new degrees of freedom are a choice of gauge on the boundary, transformations of which are generated by the normal component of the nonabelian electric field.
In general relativity the new degrees of freedom are the location of a codimension-2 surface and a choice of conformal normal frame.
These degrees of freedom transform under a group of \emph{surface symmetries}, consisting of diffeomorphisms of the codimension-2 boundary, and position-dependent linear deformations of its normal plane.
We find the observables which generate these symmetries, consisting of the conformal normal metric and curvature of the normal connection.
We discuss the implications for the problem of defining entanglement entropy in quantum gravity. Our work suggests that the Bekenstein-Hawking entropy may arise from the different ways of gluing together two partial Cauchy surfaces at a cross-section of the horizon.
}

\maketitle

\section{Introduction}

Entanglement plays an increasingly prominent role in fundamental physics.
It has long been recognized that the entanglement naturally present in the vacuum could account for the Bekenstein-Hawking entropy of black holes \cite{Sorkin1983,Bombelli1986,Srednicki1993} (see \cite{Solodukhin2011} for a review).
In fact it is conjectured that the scaling of entanglement entropy with area is a universal feature of near-vacuum states in quantum gravity \cite{Bianchi:2012ev}. 
More ambitiously, it has been proposed that the geometry of spacetime \cite{VanRaamsdonk2010} and its topology \cite{Maldacena2013} may emerge from entanglement.
Moreover, if the Bekenstein-Hawking entropy really is entanglement entropy, then all regions of space should possess an entropy and therefore be subject to the laws of thermodynamics.
Applying the laws of thermodynamics to local Rindler horizons \cite{Jacobson1995}, or to small spherical regions of spacetime \cite{Jacobson:2015hqa}, one can derive the Einstein equation as an equation of state.
Together these results point toward idea that entanglement, an essentially quantum phenomenon, is deeply connected to classical spacetime geometry and its dynamics.

Despite the central role that entanglement is conjectured to play in quantum gravity, there is so far no sharp definition of entanglement between regions of space that would apply in quantum gravity.
In order to define entanglement entropy, we require that for any Cauchy surface divided into two disjoint regions $\Sigma$ and $\overline \Sigma$, there is a tensor product factorization of the total Hilbert space
\begin{equation} \label{tensorproduct}
\mathcal{H} = \mathcal{H}_\Sigma \otimes \mathcal{H}_{\overline \Sigma}.
\end{equation}
We can then define for any density matrix $\rho$, a reduced density matrix $\rho_\Sigma := \Tr_{\overline \Sigma} \rho$ and an entanglement entropy
\begin{equation}
S(\rho_\Sigma) = - \Tr \rho_\Sigma \log \rho_\Sigma.
\end{equation}
We do not expect the Hilbert space of quantum gravity to be naturally equipped with a tensor product structure such as \eqref{tensorproduct}.
An essential property of the tensor product is that operators within the different factors commute: $[A \otimes I, I \otimes B] = 0$ for any operators $A$ and $B$.
There are strong arguments that such locally supported commuting operators simply cannot exist in quantum gravity \cite{Giddings:2005id,Giddings:2015lla}.
The basic obstruction is diffeomorphism invariance: physical operators in quantum gravity should be diffeomorphism invariant, but the action of a diffeomorphism is to permute the points of spacetime. 
Thus an invariant operator with support at any spacetime point must also have support on every other spacetime point.
Whatever notion of local subsystems survives in quantum gravity must clearly be a generalization of \eqref{tensorproduct}. 
The goal of this paper is to unravel the mechanism behind such a generalisation and the basis for its construction.

The absence of tensor factorisation is expected for any theory with gauge symmetry.
In a gauge theory the initial data must obey constraints, which take the form of differential equations on each time slice.
In Yang-Mills theory the constraint is Gauss' law, and in gravity it is the diffeomorphism constraints (often further divided into a spatial diffeomorphism constraint and a timelike Hamiltonian constraint).
Because of these constraint equations, initial data in the two regions $\Sigma$ and $\overline \Sigma$ cannot be specified independently.
This fact is reflected at the quantum level by the failure of the factorisation \eqref{tensorproduct}.

Before proceeding to describe our construction, we discuss some alternative perspectives on the problem of entanglement entropy and localized subsystems in field theory.
The most straightforward way to define entanglement entropy for a field theory is to regularize on a lattice. 
There each site carries a Hilbert space, which could be for example that of a harmonic oscillator, or a finite-dimensional system.
Then the Hilbert space of each region is simply the tensor product of all the sites inside the region, and the factorisation structure \eqref{tensorproduct} is manifest.
One can therefore define entanglement entropy for any quantum field theory that arises as a continuum limit of such a lattice system.
Although the entanglement entropy defined in this way is divergent and depends on the precise lattice realization of the theory, certain universal quantities can be extracted from it.
One might hope to define entanglement entropy in quantum gravity via such a bottom-up approach.
Of course, this requires gravity to emerge from a theory with a local tensor product structure, as in string-net models \cite{Gu:2009jh}.
It has recently been suggested that the bulk gravity theory in AdS/CFT must have an ultraviolet completion that does factor in this way \cite{Harlow:2015lma}.

In such a lattice model, gauge invariance is not fundamental.
Instead, it emerges only within a low-energy sector of the Hilbert space.
This low-energy sector does not admit a factorization into local subsystems; only the full ultraviolet completion does.
Entanglement entropy of the low-energy gauge-invariant states is then defined by embedding them into a larger Hilbert space which does factorize.
We will see that a similar embedding can be defined directly in the gauge theory, without specifying a lattice regularization.

An alternative approach is to identify a region of spacetime by its algebra of local observables, as is standard in the algebraic approach to QFT \cite{Haag1992}.
One can define the entropy of a region to be the entropy of its observable algebra, as advocated in Ref.~\cite{Casini:2013rba}.
This approach to subsystems is problematic for several reasons \cite{Giddings:2015lla}.
First, in quantum field theory the algebra of observables associated to a spacetime region is a type III von Neumann algebra \cite{Yngvason2005}. 
Unlike a type I von Neumann algebra, a type III von Neumann algebra is not associated with a factorization of the Hilbert space.
Since the type III property depends on the ultraviolet behavior of quantum field theory, it is not clear whether this property will persist in quantum gravity.
A much more serious issue is the absence of compactly supported gauge-invariant observables in gravity \cite{Giddings:2005id}.
Thus it would seem that the algebraic approach, which is based on algebras of compactly supported gauge-invariant observables, does not know how to assign an entropy to systems without local gauge-invariant operators.

Instead we will follow an approach that has been called the extended Hilbert space construction \cite{Buividovich:2008gq,Donnelly2011,Ghosh:2015iwa,Aoki:2015bsa,Soni:2015yga}.
We can divide the constraints of the theory into three classes: those with support in $\Sigma$, those with support in $\overline \Sigma$, and those with nontrivial support on the boundary.
To the region $\Sigma$ we assign an extended Hilbert space $\mathcal{H}_\Sigma$, consisting of states that satisfy the gauge condition away from the entangling surface $S = \partial \Sigma$, i.e. which are annihilated by the constraints with support only inside $\Sigma$.
These states of the extended Hilbert spaces $\mathcal{H}_\Sigma$ and $\mathcal{H}_{\overline{\Sigma}}$ generally transform nontrivially under transformations with nontrivial support on the boundary, so the Hilbert space carries a nontrivial representation of the boundary gauge group.
We can then identify the physical Hilbert space as a subspace of the tensor product 
\begin{equation} \label{embedding}
\mathcal{H} \subset \mathcal{H}_\Sigma \otimes \mathcal{H}_{\overline \Sigma}.
\end{equation}
The space $\mathcal{H}$ consists of those states that are invariant under boundary gauge transformations.
Equivalently, it is the space of states annihilated by the constraints with nontrivial support on the boundary.

The extended Hilbert space construction can be defined precisely in Yang-Mills theory on a lattice \cite{Donnelly2011} and in two dimensions \cite{Donnelly:2014gva}.
Here the extended Hilbert space $\mathcal{H}_\Sigma$ consists of Wilson loops (or more generally, generalized spin network states \cite{Donnelly2011}) which are allowed to have open ends on the boundary $S$.
This is in a sense the minimal extension of the Hilbert space: in order for $\mathcal{H}$ to contain gauge-invariant operators such as a Wilson loop in a representation $R$, the space $\mathcal{H}_\Sigma$ has to contain a state that looks like a Wilson loop in the interior of $\Sigma$, but which terminates on $S$.
The endpoint of the Wilson line acts as a surface charge transforming in the representation $R$.
The generators of this gauge transformation are the nonabelian electric fields along links that cross the boundary.
The subspace $\mathcal{H}$ consists of the gauge-invariant states for which the surface charges associated on one side of the boundary match up with the surface charges on the other side to form singlets.
Equivalently, the normal component of the electric field must match across the boundary.

The result of this construction is that the gauge symmetry of the boundary is promoted to a physical symmetry on $\mathcal{H}_\Sigma$.
The entanglement entropy that results from the embedding \eqref{embedding} includes a contribution from the surface charge degrees of freedom.
While it may seem like overcounting to count gauge-related states in the entanglement entropy, multiple lines of evidence point to this being the physically relevant definition of entanglement entropy. 
For example, in Ref.~\cite{Donnelly:2014gva} it is shown that such states are necessary for consistency of the thermal description of the de Sitter vacuum.
Additional evidence for this is provided by the calculation in Ref.~\cite{Lewkowycz2013}, where the entanglement entropy of a quark-antiquark pair was found to contain a contribution from surface charges where the nonabelian electric string pierces the entangling surface.
In abelian gauge theory, the surface charges are needed to provide agreement with Euclidean methods, and to obtain agreement between the conformal anomaly and the universal logarithmic term in the entanglement entropy \cite{Donnelly:2014fua,Donnelly:2015hxa}.
Thus the surface charge degrees of freedom are an essential component of the horizon entropy.

Our goal is twofold.
First, we want to revisit and improve the previous construction of the extended Hilbert space in gauge theory by analysing its classical analog, the extended phase space. 
Second, we would like to define an analogous extended phase space for gravity. 

We first show how to assign to each region of space $\Sigma$ an extended phase space ${\cal P}_\Sigma$, which contains additional degrees of freedom at the boundary.
This extended phase space is the classical analog of the extended Hilbert space.
Our analysis shows that gauge invariance dictates what type of additional degrees of freedom are needed and their commutation relations. 
These additional degrees of freedom are designed to make the phase space invariant under all gauge transformations of the original system, whether or not they vanish at the boundary. In addition to the gauge symmetries extended to the boundary, we discover that 
the boundary degrees of freedom transform covariantly under a new group of physical symmetries, which we call \emph{surface symmetries}, that are implemented as Hamiltonian transformations on the extended phase space. 
The appearance of a new type of boundary symmetry group $G_S$ where $S=\partial \Sigma$ as a result of gauge invariance is one of the key results of this paper.
We find, quite remarkably, that the boundary symmetry group commutes with the gauge transformations; this implies that the charges generating $G_S$ are gauge-invariant boundary observables. 
The second goal of this paper is to identify these charges for Yang-Mills and gravity.
 At the quantum level this implies that the extended Hilbert space ${\cal H}_\Sigma$ carries a representation of $G_S$.

We can  use this Hamiltonian group action to define the the phase space associated to a union of regions $\Sigma \cup \overline \Sigma$. It is given by a {\it fusion} product 
\beq 
{\cal P}_{ \Sigma \cup \overline \Sigma} = {\cal P}_\Sigma \times_{G_S}{\cal P}_{ \overline \Sigma},
\eeq where we take the Cartesian product of the phase spaces (analogous to the tensor product of Hilbert spaces) and mod out by the action of the surface symmetry group.
At the quantum level this construction is implemented by an {\it entangling} product 
\beq 
\mathcal{H}_{ \Sigma \cup \overline \Sigma} = \mathcal{H}_\Sigma \otimes_{G_S} \mathcal{H}_{ \overline \Sigma},
\eeq
where the entangling product $\otimes_{G_S}$ is a tensor product of modules:
${H}_\Sigma$ acts as a right module for $G_S$ and ${H}_{\overline \Sigma}$ acts as left module for $G_S$ and we make the identification $ ( g \ket{\psi} ; g \ket{\overline{\psi}} ) \simeq ( \ket{\psi} ; \ket{\overline{\psi}})$, for $g\in G_S$.
In other words the entangling product restricts to states that are singlets under the diagonal action of surface symmetries.

We begin by constructing an extended phase space for Yang-Mills theory with a general gauge group $G$ in section \ref{section:ym}.
We follow the covariant canonical formalism, which has been developed by many authors, and presented e.g. in \cite{Crnkovic:1987tz}.
The key step is to include the choice of gauge as a dynamical variable in the phase space,\footnote{This procedure is similar to that followed in Ref.~\cite{Regge1974}, who introduced an asymptotic coordinate system as additional dynamical variables to the phase space of asymptotically flat gravity in order to have a phase space carrying a representation of the Poincar\"e group.}
In the mathematical language this is a choice of bundle trivialization, but locally it is simply a $G$-valued function $\varphi: S\to G$.
Following the standard algorithm we define a presymplectic potential form $\theta$, which can be integrated over a Cauchy surface $\Sigma$ to define the presymplectic potential of the theory.
If the Cauchy surface $\Sigma$ has a boundary, the presymplectic potential obtained from this procedure is not gauge-invariant.
We show how it can be made invariant by adding a boundary term that depends on $\varphi$ and its conjugate momentum: the non-abelian surface charge $E$, which is a Lie algebra valued density. 
When this is done we obtain a presymplectic potential that is fully invariant under gauge transformations, including those which act at the boundary.

The introduction of the additional variable $\varphi$ allows us to introduce a new symmetry of the system, which rotates $\varphi$ without changing the other fields.
Thus we distinguish between two different types of transformations:
\begin{align}
\textbf{Gauge symmetry:}& &\delta_\alpha \varphi &= \varphi \alpha, & \delta_\alpha A &= \rd_A \alpha \\
\textbf{Surface symmetry:}& &\Delta_\alpha \varphi &= -\alpha \varphi, & \Delta_\alpha A &= 0
\end{align}
The former is pure gauge, acts on the right of the trivialisation, and has vanishing generator $H[\alpha]$.
This generator can be expressed as a bulk term proportional to the Gauss constraint, and a boundary term that enforces the identification of the normal electric field with the surface charge (see section \ref{subsection:ym-gluing}).
The latter is a symmetry, acts on the left of the trivialisation, and its generator $E[\alpha]$ is nonvanishing and is given by the the non-abelian surface charge.
The vanishing of the boundary component of the gauge generator $H[\alpha]$ can be understood as a boundary condition that identifies the non-abelian surface charge density $E$ with the normal component  of the electric field in the frame $\varphi$. Explicitly this reads 
\beq \label{dressedE}
n^a s^b F_{ab} =(\varphi^{-1} E \varphi) ,
\eeq
$n^a$ is the unit time normal to the slice $\Sigma$ and  $s^b$ is the unit normal of $S$ within $\Sigma$, $F$ is the Yang-Mills curvature. 
These normal electric fields therefore generate an algebra of non-abelian surface charges.
The Poisson brackets of this extended phase space coincide with the commutators of the extended Hilbert space of lattice gauge theory; we have thus succeeded in constructing a continuum version of the lattice construction \cite{Donnelly2011}.

The most significant advantage of our covariant formalism is that it extends naturally to general relativity, as we show in section \ref{section:gr}.
Here in addition to a metric $g_{ab}$, we introduce a variable $X^a : \R^D \to M $ that specifies the coordinate location of the entangling surface $S = X(s)$, where $s$ is a fixed $(D-2)$-dimensional closed surface in $\mathbb{R}^D$.
$X^a$ plays a role analogous to  $\varphi$ which specifies the choice of boundary gauge in Yang-Mills.
The non-temporal variable $X^a$ drops out of the presymplectic potential when integrated over a manifold without boundary. As shown in \cite{Freidel:2013jfa} the temporal component $\delta T$ does not in general, unless we restrict to hypersurface normal deformations.
On a surface $\Sigma$ with boundary, we show that the addition of an $X^a$-dependent boundary term makes the presymplectic potential diffeomorphism-invariant, even for diffeomorphisms supported on the boundary.

The additional phase space variable $X^a$ again leads to physical symmetries of the extended phase space.
Thus again we distinguish between two types of transformations:
\begin{align}
\textbf{Gauge symmetry:}& &\delta_V X^a &= -V^a(X), & \delta_V g_{ab} &= \mathcal{L}_V g_{ab} \\
\textbf{Surface symmetry:}& &\Delta_w X^a &= (\pa_b X^a) w^b , & \Delta_w g_{ab} &= 0
\end{align}
$V$ is a vector field on $M$ that labels infinitesimal diffeomorphism of spacetime: these are gauge. 
$w$ is a vector field on $\mathbb{R}^D$ that generates an infinitesimal change of the surface $S$: these are the surface symmetries.
The gravitational case is more subtle than Yang-Mills because of the different ways that a transformation of $X^a$ can act in a neighbourhood of the codimension-2 surface $S$.
Thus we distinguish between three classes of surface transformations:
\begin{itemize}
\item \textbf{Surface boosts} linearly transform the normal plane of $S$ in a position-dependent way.
\item \textbf{Surface diffeomorphisms} are reparametrizations of $S$ that move $S$ tangent to itself.
\item \textbf{Surface translations} move the surface $S$ in the normal direction.
\end{itemize}
Previous works \cite{Balachandran1994,Balachandran1995} and especially \cite{Carlip1994,Carlip1998,Carlip1999} (see \cite{Carlip2011} for a more recent overview) have focused on the surface translations.
In order to implement the surface translations as canonical transformations, boundary conditions are usually imposed at the surface $S$, and the correct way to choose those boundary conditions remains somewhat mysterious.
Here we focus instead on the first two classes of transformations, which we call the \emph{surface-preserving symmetries}. These are the boundary symmetries that preserve the location of the surface as a whole.  
We find canonical generators  for these symmetries, without the need to specify the location of the surface or any associated boundary conditions.
However it is clear that the full story must include both classes of transformations, especially if we are to consider dynamically evolving entangling surfaces.
While we hope that our formalism will lend insight to the quantization of the surface translations, we leave this task for future work.

The central result of this paper is twofold.
First, we identify the surface-preserving symmetry group which is given by the semi-direct product 
\beq
G_S =\text{Diff}(S) \ltimes \slr^S
\eeq 
of diffeomorphisms of the entangling surface, times the algebra of surface boosts represented as group valued maps  $S\to \slr$. 
We also show that the action of this symmetry group is Hamiltonian on the extended phase space of gravity and identify the Hamiltonian generators in terms of intrinsic and extrinsic geometry of $S$. 
These Hamiltonian generators play the role of gravitational surface charges for the entangling surface $S$. They are the analog of the dressed normal electric field \eqref{dressedE} for Yang-Mills.
Remarkably they can be described explicitly as follows.

To evaluate the Hamiltonian in terms of components of the metric we introduce a $2+(D-2)$ decomposition and adapted coordinates $(x^i,\sigma^\mu)$.
 $\sigma^\mu$ are coordinates on $S$, and  $(x^{i})_{i=0,1}$ are  coordinates in the two normal directions.
In these coordinates we can parametrize the metric as
\be
\rd s^{2} = h_{ij}\rd x^{i}\rd x^{j} + q_{\mu \nu}\left(\rd \sigma^{\mu} - A^{\mu}_{i}\rd x^{i}\right)
\left(\rd \sigma^\nu - A^\nu_j\rd x^{j}\right).
\ee
Here $q_{\mu \nu}$ is the induced metric on $S$, $h_{ij}$ a generalized lapse which defines the normal geometry and $A_i^\mu$ a generalized shift.
$A_i^\mu$ can be viewed as a connection on the normal bundle of $S$. 
The generators of diffeomorphisms and $\slr$ symmetries are integrals of surface charge densities, given by the normal curvature and the conformal normal geometry, respectively:
\beq
{\sqrt{q}} \left(\partial_{0}A_1^\mu-\partial_1A_0^\mu +[A_{0}, A_1]^\mu\right)/{\sqrt{h}},\qquad 
- \sqrt{q} \, h_{jk}\epsilon^{ki}/{\sqrt{h}}.
\eeq
Here $\sqrt{q}$ is the volume form on $S$, while $\sqrt{h}$ is the square root of the determinant of the normal metric, and $\epsilon^{ij}$ is the two-dimensional Levi-Civita symbol.

We see that the algebra of surface boosts is generated by conformal components of the metric normal to $S$ and form a local $\slr$ algebra whose Casimir is the area element of the surface $S$.
The diffeomorphisms are generated by the curvature of the normal connection, which vanishes if the normal planes are integrable. 
This is our main result.
This result, along with the identification of the fusion product of phase space and the entangling product, form the basis of understanding gauge systems with boundaries.

While quantization is beyond the scope of this work, our approach shows that the Hilbert space associated with $\Sigma$ must carry a unitary representation of the boundary symmetry group $G_S$. In section \ref{section:quantization} we mention some general results in representation theory which may be useful.
A key point is the identification of the area element of the surface with the Casimir of $\slr$.
The representation theory of this algebra therefore singles out the area of the entangling surface $S$ as the physical quantity that labels the dimension of the irreducible representations, and area preserving diffeomorphisms as a key subgroup to represent unitarily.
We conclude in section \ref{section:discussion} with many areas for future work.
Some technical details of our formalism have been exiled to two appendices.

\section{Yang-Mills symplectic potential} \label{section:ym}

We first construct an extended phase space for bounded regions in Yang-Mills theory.
In addition to being of interest in its own right, Yang-Mills theory provides a useful warm-up exercise for gravity and allows us to review some aspects of the covariant canonical formalism.
Moreover, it will allow us to see that the method we have developed is a continuum generalization of the extended Hilbert space approach to lattice Yang-Mills theory \cite{Buividovich:2008gq,Donnelly2011,Ghosh:2015iwa,Aoki:2015bsa,Soni:2015yga}.

Consider Yang-Mills theory with gauge group $G$.
The field variable is a connection $A$, which can be represented in a trivialization as a $\mathfrak{g}$-valued 1-form.
The Lagrangian density is given by the $D$-form
\begin{equation}
L = -\frac{1}{2} \Tr[ F \wedge \star F] .
\end{equation}
where $\star$ denotes the $D$-dimensional Hodge duality operation\footnote{Our convention is that $(\alpha \wedge \star \beta) = g(\alpha,\beta) \epsilon$ where $\epsilon$ is the volume form.} and $F=\rd A + \frac12 [A,A]$ is the curvature two-form.

The first step of deriving the canonical formalism is to find the symplectic potential of the theory, which we do following the method of Refs.~\cite{Kijowski:1976ze,Crnkovic:1987tz, Crnkovic1987, Gawedzki:1990jc, Lee:1990nz}.
The space of classical solutions of a Lagrangian system is naturally equipped with a presymplectic structure. 
The symplectic potential itself is obtained from an on-shell variation of the action.
We introduce a differential $\delta$ on field space which satisfies Leibniz rule and squares to zero: $\delta^{2}=0$. 
It acts on the space of solutions of the Lagrangian system mapping infinitesimally solutions to solutions and it is such that the product of field differentials anti commute $\delta A \delta B= -\delta B\delta A$ for $A,B$ two scalar fields.
Following the conventions of Ref.~\cite{Crnkovic1987} we do not include an explicit wedge product between the 1-forms $\delta_g$; we reserve the notation $\wedge$ for spacetime forms and keep the antisymmetrization of $\delta$ implicit.
This field space differential $\delta$ should not be confused with the spacetime differential denoted $\rd$.
A general field is a $(p,q)$ form where $p$ denotes the degree of the form in field space and $q$ the degree as a spacetime form. 

We begin by computing the symplectic potential current density $\theta$ by varying the Lagrangian,
\begin{equation}
\delta L[A] = \Tr[  \delta A \wedge (\rd_A {\star F}) ] + \rd \theta[A,\delta A].
\end{equation}
The first term gives the equation of motion, $\rd_A (\star F) := \rd(\star F) + [A,{\star F}] = 0$. 
The second term yields the symplectic potential current density:
\be
\theta[A, \delta A] = -\Tr[  \delta A\wedge \star F ].
\ee
$\theta$ depends on the field $A$ and is linear in the variation $\delta A$; it is a $(1,D-1)$-form, a one-form in field space in the sense of Ref.~\cite{Crnkovic1987} and a $(D-1)$-form in spacetime.
To obtain the presymplectic potential of the theory, we integrate $\theta$ over a Cauchy surface.
The expression $\delta A$ is analogous to the finite-dimensional expression $\rd q$ where $q$ is a canonical coordinate, and the integral of the current $\theta$ is analogous to the finite-dimensional symplectic potential $\sum_i p_i \rd q^i$.

It is important to note that $\theta$ depends explicitly on $A$, and is not gauge-invariant.
This is potentially a problem, because we would like to define $\theta$ as a one-form on the gauge orbits, and this is only possible if $\theta$ is gauge-invariant.
As we will see this poses no problem when $\theta$ is integrated over a closed surface $\Sigma$ without boundary, but when $\Sigma$ has a boundary the symplectic structure does change nontrivially under gauge transformations.
We will then show how to correct for this non-invariance by addition of a boundary term to the symplectic structure.

Let $g^*$ denote the gauge transformation by $g$, which acts on $A$ and $F$ as
\bea
A & \rightarrow & g^*(A) = g A g^{-1} - (\rd g) g^{-1}, \\
F & \rightarrow & g^*(F) = g F g^{-1}.
\eea
Now consider how $\theta$ changes under this gauge transformation.
After transformation the new symplectic potential current density is given by
\bea  
\theta[g^*(A), \delta g^*(A)] &=& -\Tr [ (\delta (g A g^{-1}) - \delta(\rd g g^{-1}))\wedge  \star   g F  g^{-1}] \nonumber \\
&=& -\Tr[  \left(\delta A - \rd (g^{-1} \delta g) - [A, g^{-1}\delta g] \right) \wedge \star F ]. \label{thetagA}
\eea
We now introduce some notation that will clarify this calculation and generalize more readily to the gravitational case.
The quantity $\delta g$ is a somewhat unnatural object: it is an element of the tangent space to $G$ at the point $g$. 
A more natural object is obtained by translating it back to a neighbourhood of the identity:
\begin{equation}
\delta_g := g^{-1} \delta g.
\end{equation}
$\delta_g$ is an element of the Lie algebra, and the transformation of the presymplectic potential form \eqref{thetagA} can be expressed in terms of the gauge covariant derivative of $\delta_g$:
\be
\rd_A(\delta_g) = \rd (g^{-1} \delta g) + [A, g^{-1} \delta g].
\ee
Thus we have
\be \label{thetadeltaG}
\theta[g^* A, \delta g^* A] = -\Tr[ (\delta A - \rd_A (\delta_g) )\wedge \star F  ].
\ee
The presence of the second term in Eq.~\eqref{thetadeltaG} shows precisely the degree to which the symplectic potential current density is not gauge invariant.

The reason that $\theta$ is not gauge invariant is that the variation $\delta$ does not commute with the gauge transformation $g^*$.
Instead the difference can be expressed as an infinitesimal gauge transformation by $\delta_g$:
\begin{equation} \label{deltaGA}
\delta g^* (A) = g^* (\delta A - \rd_A \delta_g ).
\end{equation}
Note that $\delta A$ being a difference of two connections transforms as $g^* (\delta A) = g \, \delta A \, g^{-1}$.
Similarly, for the field strength $F$ we have
\begin{equation} \label{deltaGF}
\delta g^* (F) = g^* (\delta F + [\delta_g, F]),
\end{equation}
which is again the transformation rule for $F$ under an infinitesimal gauge transformation by $\delta_g$.

Later we will need the variation of $\delta_g$ itself, which is given by
\begin{equation} \label{deltadeltag}
\delta (\delta_g) = -\tfrac12 [\delta_g, \delta_g].
\end{equation}
Recall that the two $\delta$'s in this formula correspond to two different (anticommuting) variations.

To obtain the presymplectic potential $\Theta$ of the theory, we integrate the symplectic potential current over a Cauchy surface $\Sigma$:
\begin{equation} \label{YMtheta}
\Theta_\Sigma[A, \delta A] = \int_\Sigma \theta[A, \delta A ]= -\int_\Sigma \Tr\left[  \delta A \wedge\star F\right].
\end{equation}
This is not manifestly gauge invariant because of the nontrivial transformation law for $\theta$ in Eq.~\eqref{thetadeltaG}.
But the latter term can be integrated by parts, using the equation of motion $\rd_A{\star }F \hateq 0$, with $\hateq$ denoting equality on shell.
We find that the presymplectic potential changes by a boundary term,
\begin{equation} \label{boundary-term}
\Theta_\Sigma[g^*A, \delta g^* A] \hateq \Theta_\Sigma[A, \delta A] + \oint_S \Tr[\star F \delta_g].
\end{equation}
When $\theta$ is integrated over a Cauchy surface without boundary, or when the boundary conditions are such that the boundary term in \eqref{boundary-term} vanishes, the presymplectic potential is gauge invariant.
However we are interested in associated a phase space with a region $\Sigma$ whose boundary $S$ is the entangling surface.
In this case we do not wish to impose any particular boundary conditions at $S$, since the entangling surface is not a physically distinguished location as far as the gauge field is concerned.

In order to define a gauge-invariant symplectic potential we need to add to $\Theta_\Sigma$ a term whose gauge transformation compensates that in Eq.~\eqref{boundary-term}.
To construct such a potential we recall that a connection on a bundle is not specified just by the 1-form $A$ but also by a local trivialization $\varphi$, which we may think of, at least locally, as a $G$-valued function $\varphi:S \to G$.
Under a gauge transformation $\varphi$ transforms as
$g^*(\varphi) = \varphi g^{-1}$.
We can now augment the symplectic structure with the term
\begin{equation}
\Theta_S[A, \varphi, \delta \varphi] = \oint_S \Tr[\star F \delta_\varphi]
\end{equation}
where analogous to $\delta_g$ we define $\delta_\varphi := \varphi^{-1} \delta \varphi$.
Under a gauge transformation $\delta_\varphi$ transforms as
\begin{align} \label{deltagstarphi}
\delta_{g^*(\varphi)} = g^*( \delta_\varphi -\delta_g)
\end{align} which leads to the gauge transformation law for $\Theta_S$
\begin{equation} \label{boundary-term2}
\Theta_S[g^* A, g^*\varphi, \delta g^* \varphi] = \Theta_S[F, \varphi, \delta \varphi] - \oint_S \Tr[\star F \delta_g].
\end{equation}
The second term in \eqref{boundary-term2} is exactly what is needed to cancel the second term in \eqref{boundary-term}.
Combining these two terms we obtain the presymplectic potential
\bea
\Theta[A, \delta A, \varphi, \delta \varphi] &=& \Theta_\Sigma[A, \delta A] + \Theta_S[A, \varphi, \delta \varphi] \\
&=& -\int_\Sigma \Tr[ \delta A \wedge \star F ] + \oint_S \Tr[\star F \delta_\varphi].
\eea 
This presymplectic potential is gauge invariant for any spacelike surface $\Sigma$, even for gauge transformations with support on the boundary.

\subsection{Gauge invariance} 
The above presymplectic potential satisfies, by construction, the gauge invariance property
\be
\Theta[g^* A, \delta g^*A, g^*\varphi, \delta g^* \varphi] = \Theta[A, \delta A, \varphi, \delta \varphi]\label{gaugeinv}.
\ee
Note that this depends not only on the gauge field $A$ everywhere on $\Sigma$ (including the boundary $S$), but also on the trivialization $\varphi$ on $S$.
Because $\Theta$ is gauge-invariant, we can pull it back to define a symplectic potential on the space gauge orbits, which is the true physical phase space of the theory.
This means that for any $\alpha$ valued in the Lie algebra $\mathfrak{g}$, the infinitesimal gauge transformation
\be \label{gaugetrans}
\delta_\alpha \varphi = \varphi \alpha, \qquad
\delta_\alpha A = \rd_A \alpha,
\qquad \delta_\alpha F = [F,\alpha].
\ee 
is a null direction of the total symplectic form, for any $\alpha$ including those with support on the boundary.
We will now show explicitly that the canonical generator for this transformation vanishes on-shell. 
To find the generators of these transformations, we take the variation of the symplectic potential to obtain  the presymplectic variational 2-form $\Omega =\delta \Theta$.
This presymplectic form is the sum of a bulk term $\Omega_\Sigma$ and a boundary term $\Omega_S$, where the latter depends on the boundary degree of freedom $\varphi$:
\bea
\Omega_\Sigma &=&  \int_\Sigma \Tr[ \delta A\wedge \star \delta F ], \\
\Omega_S &=&  \oint_S \Tr[\star \delta F \delta_\varphi - \star F \delta_\varphi \delta_\varphi].
\eea

We are now interested in constructing the generators of gauge transformations associated with the bulk and boundary symplectic structure in order to determine the Poisson brackets between boundary observables. 
Recall that a Hamiltonian $H$ determines a variational vector field $\delta_H$ via 
\begin{equation} \label{poisson}
\Big\{H , f \Big\} = \delta_H f.
\end{equation}
Since we have already introduced a field differential $\delta$ it is natural to also introduce a field ``interior product'': given any infinitesimal field transformation $\delta_\alpha A$, we denote $I_{\delta_\alpha}$ the inner product along this variation that maps $n$-form in field space to $(n-1)$-forms.
$I_{\delta_\alpha}$ acts trivially on functionals of $A$ (field space $0$-forms), satisfies the graded Leibniz rule (see appendix \ref{appendix:exterior}) and is defined on field one-forms by
\be
I_{\delta_\alpha} \delta A := {\delta_\alpha}A.
\ee
The Poisson bracket \eqref{poisson} can then be expressed in terms of the interior product and the symplectic structure $\Omega$ via   
\be
I_{\delta_F}I_{\delta_H}\Omega=[I_{\delta_H}\Omega] (\delta_F) ={\Omega}(\delta_H,\delta_F) = \Big\{F,H\Big\}.
\ee 
where $I$ is the interior product. 
Thus the Hamiltonian $H$ and the variational vector field it generates are related by
\be \label{HdeltaH}
I_{\delta_{H}}\Omega =\Omega(\delta_H,\cdot)= \delta H.
\ee
We can use the relation \eqref{HdeltaH} to find the generator of gauge transformations.
Let $\alpha$ be a Lie-algebra-valued function on $\Sigma$ and $\delta_\alpha$ denote the gauge transformation (\ref{gaugetrans}).
We introduce the bulk and boundary Hamiltonian generators $H_\Sigma[\alpha]$ and $H_S[\alpha]$ respectively such that
$
I_{\delta_\alpha}\Omega_\Sigma = \delta H_\Sigma[\alpha]$,  and $ 
I_{\delta_\alpha}\Omega_S = \delta H_S[\alpha]$.
One finds that they are given by
\bea
H_\Sigma[\alpha]= \int_\Sigma \Tr [ \rd_A \alpha \wedge \star F ],\qquad
H_S[\alpha]=- \int_S \Tr [\star F \alpha].
\eea
For instance we can explicitly check the boundary variation 
\bea
\delta H_S[\alpha] = -\oint_S \star \Tr [ \delta F\, \alpha ] &=&-
 \oint_S \star \Tr [  \delta F (\varphi^{-1} \delta_\alpha \varphi)+ (-\delta_\alpha F + [F, \varphi^{-1} \delta_\alpha \varphi]) \varphi^{-1} \delta \varphi]=  I_{\delta_\alpha} \Omega_S. \nonumber
\eea
We have used that the second term vanishes since $
(\varphi^{-1} \delta_\alpha \varphi) =\alpha$.
We can now check that the full generator $H[\alpha] = H_\Sigma[\alpha] + H_S[\alpha]$ vanishes on-shell:
\be
H_\Sigma[\alpha] + H_S[\alpha] =- \int_\Sigma \Tr [\alpha (\rd_A \star F)  ] \hateq 0.
\ee
This is the expression of the gauge invariance of our formulation.

\subsection{Formal derivation of gauge invariance}\label{formalgauge}

Since we would like to generalize our results to the case of gravity, and since direct computations in gravity such as the one done in the previous section are very cumbersome, it is convenient to develop a formal but simpler proof of the gauge invariance of the presymplectic potential. 
Such a proof will then generalize easily to the gravity case and also be conceptually clearer. 
Moreover, this proof elucidates the key mechanisms behind the restoration of gauge invariance by the addition of boundary degrees of freedom.

Using the field interior product we can define a field Lie derivative\footnote{$L$ and its properties are described in more detail in appendix \ref{appendix:exterior}.}
\be
L_{\delta_\alpha} := \delta I_{\delta_\alpha} + I_{\delta_\alpha}\delta .
\ee
We can now evaluate
\be
I_{\delta_\alpha} \Omega = L_{\delta_\alpha} \Theta - \delta(I_{\delta_\alpha}\Theta).
\ee
The symplectic potential we have constructed is invariant under gauge transformations, which was the main motivation to introduce the trivialisation degree of freedom $\varphi$. 
Indeed the infinitesimal version of \eqref{gaugeinv} when $g = 1 + \alpha + O(\alpha^2)$ reads 
\be
L_{\delta_\alpha}\Theta=0.
\ee
This shows that the  Hamiltonian is simply given by 
 $H[\alpha]=-I_{\delta_\alpha}\Theta$. 
 This can be evaluated directly
 \bea
 I_{\delta_\alpha}\Theta = -\int_\Sigma \Tr[ \delta_\alpha A \wedge \star F ] + \oint_S \Tr[\star F \alpha] 
 = \int_\Sigma \Tr[\alpha  (\rd_A \star F) ] \hateq 0.
 \eea
Here we recognize the contraction of the Yang-Mills equation of motion with the gauge parameter $\alpha$ pulled back to $\Sigma$.
 As we will see the same derivation applies in the gravity case.

\subsection{Boundary observables }

The addition of $\varphi$ to the phase space allows us to construct new gauge-invariant operators. 
These are operators that depend on the choice of trivialization on the boundary.
One is the electric field normal to the boundary, $ \varphi {\star F} \varphi^{-1}$.
Another is a Wilson line that begins and ends on the boundary,
\begin{equation}
W(\gamma) = \varphi(x_1) \; \mathcal{P} \exp \left(i \int_\gamma A \right) \; \varphi(x_2)^{-1}
\end{equation}
where $\gamma$ is a curve with endpoints $x_1$ and $x_2$ on the boundary.

Let us now focus on the gauge invariant normal electric field observable:
\be
E[\alpha] := \oint_S \Tr[  \star \varphi F \varphi^{-1} \alpha].
\ee
Its variation is given by 
\bea
\delta E[\alpha] &=& \oint_S \Tr\left[\star \left(\delta F + [\delta_\varphi, F]\right) \varphi^{-1}\alpha\varphi \right] = I_{{\Delta}_\alpha } \Omega
\eea
where we define the variational vector field $\Delta_\alpha$ such that $I_{\Delta_\alpha} \delta_\varphi  =- \varphi^{-1}\alpha \varphi$. 
$\Delta_\alpha$ acts only on the boundary trivialisation, so we have
\be \label{Deltaalpha}
\Delta_\alpha \varphi = - \alpha \varphi,\qquad \Delta_\alpha A=0.
\ee
$E[\alpha]$ therefore generates a local rotation of the boundary trivialisation. This rotation acts on the left of the trivialisation, unlike the gauge transformation which acts on the right.
This equation is equivalent to the statement that the Poisson bracket of $\{E[\alpha],\cdot\} =\Delta_\alpha$ with a boundary observable generates the transformation $\Delta_\alpha$.
Accordingly the previous equation determines the Poisson brackets between $E_\alpha$'s  via
\be \label{ymcommutators}
\{ E[\alpha], E[\beta] \} = \Delta_\alpha E[\beta] = -\oint_S \star \Tr[\beta [\alpha,\varphi F\varphi^{-1}]] = E[{[\alpha,\beta]}].
\ee 
Thus we find that the different components of $E:=\star (\varphi F\varphi^{-1})$ pulled back on $S$ do not commute, but satisfy the current algebra of the group $G$.
Expressed in terms of a set of generators, $E = E^a \tau_a$ with $[\tau_a,\tau_b] = f_{ab}{}^c \tau_c$ and $\Tr[\tau_a \tau_b] = \delta_{ab}$, we can expand \eqref{ymcommutators} to obtain the explicit commutation relations
\bea
\{  E^a(x),  E^b(y) \} &=& \delta(x - y) f^{ab}{}_{c} E^c(x),\\
\{  E^a(x), \varphi(y) \} &=& -\delta(x - y)  \tau^a \varphi(x),\\
\{ \varphi(x), \varphi(y) \} &=& 0.
\eea 
This is the current algebra of the group $G$, i.e. the algebra satisfied by nonabelian gauge charges on the surface $S$.
Thus the dressed normal electric field $E$ acts as a surface charge.

\subsection{Gauge versus symmetry }

As a summary, we see that the introduction of the boundary degree of freedom $\varphi$ has two effects. 
First, it restores the gauge invariance of the theory.
Second, it introduces a new  boundary symmetry which rotates the boundary observables $(\varphi, E)$.
The gauge transformations acts on the bulk variables and on the {\it right} of the boundary trivialisation 
\begin{equation} 
\delta_\alpha \varphi = \varphi \alpha, \qquad
\delta_\alpha A = \rd_A \alpha.
\end{equation}
Its Hamiltonian $H[\alpha]=\int_\Sigma \Tr[(\rd_A\star F)\alpha]$ vanishes on-shell.
The boundary symmetry leaves the bulk variables invariant and acts on the {\it left} of the boundary trivialisation
\be 
\Delta_\alpha \varphi = - \alpha \varphi,\qquad \Delta_\alpha A=0.
\ee 
with Hamiltonian $E[\alpha]=\oint_S \Tr[  \star (\varphi F \varphi^{-1}) \alpha]$. 
As a result, gauge transformations commute with symmetries. 
This implies that the symmetry generator $E[\alpha]$ is gauge invariant.
At the quantum level the Hilbert space $\mathcal{H}_\Sigma$ is annihilated by the constraints $H[\alpha]$ and carries a nontrivial representation of the  symmetry algebra generated by $E[\alpha]$.

\subsection{Extended phase space and gluing} 
\label{subsection:ym-gluing}

We are now able to define precisely the extended phase space ${\cal P}_\Sigma$ associated with an open region $\Sigma$ and the gluing property of phase space associated with  two regions $\Sigma_L$, $\Sigma_R$.
A key notion in order to describe these two procedures is the concept of symplectic reduction. 
Consider a phase space ${({\cal P},\Omega)}$ that admits a group action: $G\times {\cal P} \to {\cal P}$.
This action is  Hamiltonian if it preserves the symplectic structure.
For an infinitesimal transformation $\delta_V$ this translates to the condition
\beq
L_{\delta_V} \Omega =0.
\eeq
From the Cartan definition of $L_{\delta_V}$ and the closure of $\Omega$ we conclude that $\delta I_{\delta_V}\Omega =0$, hence locally in field space we have $I_{\delta_V}\Omega = \delta H_V$. 
The action is said to be globally Hamiltonian when the identity $I_{\delta_V}\Omega = \delta H_V$ is valid globally on field space\footnote{The appearance of central charges $c$ are due to global obstructions where $ I_{\delta_V} \Omega =\delta H_V + c $. }. 
$H$ is the Hamiltonian of the group action.

Given such a Hamiltonian group action we can define the symplectic quotient which is denoted by a double quotient labelled either by the group or by the set of Hamiltonians:
\beq
{\cal P} /\!/ G = {\cal P} /\!/ \{H\}
\eeq
The procedure is well-known from the theory of constrained Hamiltonian systems \cite{Weinstein1979}. 
First one considers the constraint subspace $C_H \subset {\cal P}$ on which the constraint $H=0$ is imposed. 
$C_H$ is naturally equipped with a presymplectic structure $\Omega_H = i_H^*\Omega$ obtained by the pullback of the embedding  map $i_H: C_H \hookrightarrow {\cal P} $. 
To obtain a symplectic manifold we need to quotient out the action of the group $G$, so that the phase space is then given by $C_H/G$. 
In the more general case where we are given a family of Hamiltonians $\{ H \}$  we only need to divide out $C_H$ by the kernel $\Ker(\Omega_H)$, so that the phase space is ${\cal P} /\!/ \{H\}:= C_H/\Ker(\Omega_H)$. 
The advantage of this description is that it treats first and second class constraints simultaneously. 
The vector fields in $\Ker(\Omega_H)$ are the infinitesimal generators of a group of transformations we denote $G_H$. 
The double quotient notation refers to the fact that the phase space is obtained by a double operation: first a {\it restriction} onto the kernel $C_H$ of constraints, and then then a {\it quotient} of this set by the group $G_H$ generated by the Hamiltonian vectors preserving $C_H$. 
The group $G_H$ and the Hamiltonians $H$ therefore play a dual role in this double quotient.

We can now use the symplectic reduction to define a classical analog of the entangling product. Suppose that we have two symplectic spaces $({\cal P}_1,\Omega_1)$ and $({\cal P}_2,\Omega_2)$ and consider the constraints $H_{12} :=  H_1-H_2=0$, which generate the group of transformations $G_{12}$. 
We can then define the fusion product 
\beq
{\cal{P}}_1 \times_{G_{12}} {\cal{P}}_2 
:= \left({\cal{P}}_1 \times  {\cal{P}}_2\right)  /\!/ G_{12}.
\eeq 
This fusion product or classical entangling product identifies the two Hamiltonians $H_1 = H_2$ and divides out by their flow. 
Thus in the fusion product the phase space variables are those acting on the space in which $H_1=H_2$ that are also invariant under its action: observables $O$ such that $\{O,H_1-H_2\}=0$.

We can now describe precisely the extended phase space for Yang-Mills.
This description formalises the construction that was first given in \cite{Freidel:2015gpa}, in the context of first order gravity.
One starts with a bulk phase space $(\hat{\cal P}_{{\Sigma}},\hat{\Omega}_{\Sigma})$, whose phase space variables are the pull-back of $(\star F , A) $ on $\Sigma$ and a boundary phase space 
$(\hat{\cal P}_{S},\hat{\Omega}_{S})$ whose phase space variables are given by  the pairs  $(E , \varphi) $ on $S$. 
Here $E$ is a $(D-2)$-form on $S$ valued in the Lie algebra and $\varphi$ is a $G$-valued function on $S$.
The symplectic structures are given by 
\beq
\hat{\Omega}_{\Sigma} =\int_\Sigma \Tr( \delta A\wedge \star \delta F ),\qquad \hat{\Omega}_{S} =\oint_S \Tr( \delta E   \delta \varphi \varphi^{-1}  + E  \delta \varphi\varphi^{-1} \delta \varphi \varphi^{-1} ).
\eeq
On the entangling surface $S$ we recognise the natural symplectic structure of the group  cotangent bundle  $T^*(G^S) \simeq (G \ltimes \mathfrak{g})^S$.
The fusion of these symplectic structures is obtained by imposing the boundary condition 
\beq\label{bdyc}
H_S[\alpha]  := \oint_S \left[\left(\star  F  \right) -\varphi^{-1} E \varphi \right]\alpha =0
\eeq
for $\alpha \in \mathfrak{g}^S$.
The extended phase space associated with $\Sigma$ with boundary $\partial \Sigma =S$ is simply the fusion product
\beq
{\cal P}_{{\Sigma}} := \hat{\cal P}_{{\Sigma}} \times_{H_S}\hat{\cal P}_{{S}}.
\eeq
What is remarkable is the fact that the boundary condition (\ref{bdyc}) which determine the fusion product is in fact {\it dictated} by  gauge invariance. Indeed as we have seen the generator of gauge invariance $H[\alpha]$  contains two contributions, one from the bulk and one from the boundary. The boundary part of the Hamiltonian generator is given by (\ref{bdyc})
\beq
H[\alpha]= H_\Sigma[\alpha] + H_S[\alpha],\qquad 
 H_\Sigma[\alpha] = -\int_\Sigma \Tr[(\rd_A \star F)\alpha].
 \eeq
 The gluing constraints (\ref{bdyc}) can therefore be understood as the boundary component of the Gauss law and results from demanding that $H[\alpha]=0$ which imposes gauge invariance. 
 Note that in this language the generator of boundary symmetries $G_S$ is simply 
 \beq
E[\alpha] = \oint_S \Tr[E\alpha].
\eeq
 We can now describe precisely the gluing procedure of two regions $\Sigma_L$ and $\Sigma_R$ sharing a common boundary $S = \Sigma_L \cap \Sigma_R$. It is given by the fusion product associated with  the constraints
 \beq
E_L[\alpha] - E_R[\alpha] =0.
\eeq
This constraint generates a diagonal action of the boundary symmetry group on ${\cal P}_{\Sigma_L}$ and ${\cal P}_{\Sigma_R}$. 
The extended phase space associated with the region $\Sigma_L\cup \Sigma_R$ is simply given by the fusion product
\beq
{\cal P}_{\Sigma_L\cup \Sigma_R} := 
{\cal P}_{\Sigma_L} \times_{G_S} 
{\cal P}_{ \Sigma_R}=
\left( \hat{\cal P}_{{\Sigma_L}} \times_{H_{S_L}}\hat{\cal P}_{{S_L}} \right) \times_{G_S}
\left(
\hat{\cal P}_{{S_R}} \times_{H_{S_R}} 
\hat{\cal P}_{{\Sigma_R}} 
\right)
\eeq

We have described the procedure of extension and gluing at the classical level in terms of phase space. This procedure has a direct, and in some ways simpler, analog at the quantum level.
At the quantum level the phase space ${\cal P}_{\Sigma}$ is replaced with a Hilbert space ${\cal H}_\Sigma$. 
The quantum analog of the symplectic quotient ${\cal P}/\!/ G$ is simply the quotient ${\cal H}/G$ of the Hilbert space by the group action.
A simple quotient is enough since demanding invariance under the action of $G$ is equivalent to imposing the Hamiltonian constraint $H \ket{\psi} = 0$, since the Hamiltonian $H$ generates the action of $G$. 
That is at the quantum level we have the equality $\Ker(H)= {\cal H}/G$, which makes the symplectic reduction a conceptually  simpler operation.\footnote{When $H$ has a continuous spectrum, the kernel of $H$ may be empty, and one must use the formalism of rigged Hilbert spaces \cite{Giulini:1998rk} to define the symplectic reduction.}
Accordingly, the extended Hilbert space is simply given by 
\beq 
{\cal H}_{{\Sigma}} =
\Ker(H_S) \subset \hat{\cal H}_{{\Sigma}} \otimes \hat{\cal H}_{{S}}
\eeq
Similarly the gluing of two extended Hilbert spaces is obtained by considering the space of singlets under the boundary symmetry, that is by considering the entangling product
\beq
{\cal H}_{\Sigma_L\cup \Sigma_R} = 
{\cal H}_{\Sigma_L} \otimes_{G^S} 
{\cal H}_{ \Sigma_R}.
\eeq

\section{Gravity symplectic potential} \label{section:gr}

We now turn to general relativity.
The formalism developed in section \ref{section:ym} generalizes with some additional features.
We will see that again the symplectic potential can be made gauge-invariant by addition of a boundary symplectic potential.
Just as the boundary symplectic potential in Yang-Mills depends on the choice of gauge at the boundary, the symplectic potential of general relativity depends on a choice of coordinates describing the location of the boundary.

In gravity we have the Lagrangian density (in units where $8 \pi G = 1$)
\be
L[g] = \tfrac12 R \epsilon.
\ee
where $R$ is the Ricci scalar of the metric $g_{ab}$, $\epsilon := \frac{s_{g}}{D!} \epsilon_{a_1 \ldots a_D} \rd x^{a_1} \wedge \cdots \wedge \rd x^{a_D}= \sqrt{g} \, \rd^Dx$ is the volume form, and $s_g$ is a sign factor that is $-1$ when $g_{ab}$ is Lorentzian, and $+1$ when $g_{ab}$ is Euclidean.
The symplectic potential current density is given by
\be \label{theta}
\theta[g, \delta g] = \frac12 \nabla_b \left[\delta g^{ab} -g^{ab}\delta g \right]\epsilon_{a}
\ee
where $\epsilon_a := \frac{s_{g}}{(D-1)!} \epsilon_{a b_2 \ldots b_D} \rd x^{b_2} \wedge \cdots \wedge \rd x^{b_D}=i_{\pa_a} \epsilon $ is the $(D-1)$-form defining the Hodge duality.
Here we used the abbreviated notation $\delta g = g^{ab} \delta g_{ab}$, and $\delta g^{a b} = g^{ac} g^{bd} \delta g_{cd}$ (i.e. $\delta g^{ab}$ is the variation of the metric with the indices raised, \emph{not} the variation of the inverse metric, which is given by $\delta (g^{ab})=-\delta g^{ab}$).
This follows from the direct calculation of the variation of the Ricci tensor given by
\begin{equation}
\delta R_{ab} = 
\tfrac12 \nabla^c  \left( \nabla_a \delta g_{bc}
+\nabla_b \delta g_{ac} - \nabla_c \delta g_{ab}
\right)- \tfrac12 
\nabla_a \nabla_b \delta g
\end{equation}
In the following we will use three notions of equality: $=$ when equality holds off-shell, $\hateq$ when the vacuum equation of motion for the background field is satisfied $G_{ab} \hateq 0$ but the variation $\delta g_{ab} $ is arbitrary and $\doteq$ when the variation is also restricted to be on shell $\delta G_{ab} \doteq 0$.

We now consider how $\theta$ transforms under the gauge symmetries of general relativity, the diffeomorphism group.
Let $Y: M \to M$ be a diffeomorphism of spacetime and denote by $Y^*: T^*M \to T^* M$ the pullback under this diffeomorphism. We also denote by $Y_*: T M \to T M$ the push forward.
Since $Y$ is invertible, the pullback map can be applied to vectors by defining $Y^* =(Y^{-1})_*$ on vectors and then to tensors with both upper and lower indices  (see appendix C of \cite{Wald1984}).
In particular, the pullback of the metric is given by
\begin{equation}
(Y^*g)_{ab}(x) = \partial_a Y^A(x) \partial_b Y^B(x) g_{AB}(Y(x)).
\end{equation}

Just as in Yang-Mills, the variation $\delta$ does not commute with the pullback $Y^*$ due to terms that involve the variation of $Y$.
Instead we have the following relation:
\begin{equation} \label{deltaXstar}
\delta Y^*(T) = Y^*(\delta T + \mathcal{L}_{\delta_Y}T)
\end{equation}
where $\mathcal{L}_V$ denotes the spacetime Lie derivative along the vector field $V$ and 
where we have introduced the vector field
\be \label{deltaX}
\delta_Y^{a}(x) := (\delta Y^{a}\circ Y^{-1})(x).
\ee
Eq.~\eqref{deltaX} is the generalization of the equations \eqref{deltaGA} and \eqref{deltaGF} of Yang-Mills to the diffeomorphism group. 
The relation \eqref{deltaXstar}, which we will use repeatedly, is proved in Appendix \ref{appendix:pullback} for general tensor-valued variational forms.

Now using \eqref{deltaXstar} we see that under a diffeomorphism of the underlying fields $\theta$ transforms as:
\be \label{newtheta}
\theta[Y^{*} g, \delta Y^{*} g] = Y^* \left(\theta[g, \delta g] + \theta[g, \mathcal{L}_{\delta_Y} g]\right).
\ee
This is not simply the pullback of the symplectic potential, because of the second term in \eqref{newtheta}.
However we can evaluate it on shell:
\begin{eqnarray}
\theta[g, \mathcal{L}_{\delta_Y}g] 
&=& \frac{1}{2} \left[ \nabla_b \left(\nabla^a \delta_Y^b +  \nabla^b \delta_Y^a\right) - 2 \nabla^a \nabla_b \delta_Y^b \right] \epsilon_a \nonumber \\
&=&  \left[  \left[\nabla_b \nabla^a -  \nabla^a \nabla_b \right]\delta_Y^b + \frac{1}{2} \nabla_b \left(  \nabla^b \delta_Y^a- \nabla^a \delta_Y^b \right) \right] \epsilon_a \nonumber \\
&=& \left[ R^a_{\phantom{a}b} \delta_Y^b  + \frac{1}{2} \nabla_b (\nabla^b \delta_Y^a - \nabla^a \delta_Y^b) \right] \epsilon_a \nonumber \\
&\hateq& \frac{1}{2} \rd \star \rd g \delta_Y. \label{thetadiff}
\end{eqnarray}
In the last line we evaluate the symplectic potential on-shell, and in the hatted equality $\hateq$ we have made use of the vacuum Einstein equation $R_{ab} = 0$.
For the last term we use the identity
\be
\nabla_{b}(\nabla^{a}V^{b}- \nabla^{b}V^{a}) \epsilon_{a}=  \rd \star \rd g V 
\ee
where we have used the notation $g$ for the operator of lowering an index on a vector to obtain a $1$-form, $g V = V_{a}\rd x^{a}$,  and $\star$ is the Hodge star acting on 2-forms\footnote{ We recall that the convention for Hodge duality on a $p$-form $\alpha = \frac1{p!} \alpha_{a_1\cdots a_p} \rd x^{a_1}\wedge  \cdots \wedge \rd x^{a_p} $ is 
\beq
(\alpha \wedge \star \beta) = \frac1{p!} \alpha_{a_1\cdots a_p} \beta^{a_1\cdots a_p} \epsilon
\eeq 
where $\epsilon$ is the volume form.} which is given by
\begin{equation}
\star (\alpha_{ab} \rd x^{a} \wedge \rd x^{b}) 
= \alpha^{ab}\epsilon_{ab},\qquad \epsilon_{ab} =   \frac{1}{(D-2)!} \epsilon_{abc_1 \ldots c_{D-2}} \rd x^{c_1} \wedge \cdots \wedge \rd x^{c_{D-2}}. 
\end{equation}
Here $\epsilon_{ab} = i_{\pa_b} i_{\pa_a}\epsilon$ is the  $(D-2)$-form obtained by double contraction with the volume form. 

Note that the previous equality can be summarised by introducing a variational vector field $I_V \delta g = \mathcal{L}_V g$.
Then the equality becomes
\begin{equation}\label{IT}
I_V \theta[g,\delta g] = \theta[g,{\cal L}_V g]= \rd \pi_g[V] + \epsilon_a R^a{}_b V^b \hateq \rd \pi_g[V].
\end{equation}
This is an identity that we will repeatedly use.
Here $\pi_g[V]$ is a $(D-2)$-form which depends linearly on $V$ and defined to be 
\begin{equation}
\pi_g[V] := \tfrac{1}{2} \star \rd g V.
\end{equation}
This shows that under diffeomorphism the symplectic potential current changes by shift under $Y$ and a total derivative:
\begin{equation} \label{grboundaryterm}
\theta[Y^{*} g, \delta Y^{*} g] = Y^* \left(\theta[g, \delta g] + \rd \pi_g[\delta_Y] \right).
\end{equation}
We now consider the symplectic potential $\Theta_\Sigma$ obtained by integrating the current $\theta$ over the slice $\Sigma$ with boundary $\pa\Sigma =S$:
\begin{equation}
\Theta_\Sigma[g, \delta g] := \int_\Sigma \theta[g, \delta g].
\end{equation}
$\Theta_\Sigma$ is not invariant under diffeomorphism, instead we have
\begin{equation} \label{thetasigmayg}
\Theta_\Sigma[Y^*g, \delta Y^*g] = \int_{Y(\Sigma)} \theta[g, \delta g] +  \int_{Y(S)} \pi_g[\delta_Y].
\end{equation}
To arrive at this expression we have used the integral identity
\begin{equation}
\int_{Y(\Sigma)} \omega = \int_{\Sigma} Y^* \omega,
\end{equation}
for a differential form $\omega$.
There are two sources of diffeomorphism non-invariance in this expression.
The first is that we integrate over the surface $Y(\Sigma)$ with boundary $Y(S)$: this simply reflects that the diffeomorphism moves the coordinate location of the surface over which we integrate.
The second source of diffeomorphism non-invariance is the presence of the second term in \eqref{thetasigmayg}, which can be cancelled by addition of a boundary symplectic potential as we will now show.

As in the case of Yang-Mills, we introduce a new variable into the phase space whose transformation law will cancel the boundary term in the symplectic potential.
In gauge theory, this extra degree of freedom was a trivialization of the bundle; in a diffeomorphism-invariant theory, the natural analog is a coordinate system $X$ which we view as a mapping  $X:U \to M$ where $U \subset \R^D$ is an open set. 
$X$ is assumed to be invertible on its image. 
We assume for simplicity that $\Sigma$ can be covered with one open set and  we denote by $\sigma \subset U$ the preimage of $\Sigma$, with boundary $\pa\sigma =s$. 
Under this map we have $\Sigma=X(\sigma) $ and $S= X(s)$.
Under a diffeomorphism $Y:M \to M$, the coordinate system $X$ changes to $Y^*(X) = Y^{-1} \circ X$. We can use this coordinate system to define the symplectic potential $\Theta_{\Sigma}[g, \delta g]$ as the integral over the slice $\Sigma = X(\sigma)$. 
Under a diffeomorphism its transformation is given by
\begin{equation}
\Theta_{(Y^{-1} \circ X)(\sigma)}[Y^*g, \delta Y^*g] = \Theta_\Sigma[g, \delta g] + \int_S \pi_g[\delta_Y].
\end{equation}
Thus when integrating over the surface $\Sigma$ the only change in the symplectic structure under diffeomorphism is via the boundary term.

In order to cancel this boundary term in the transformation law of $\Theta_\Sigma$, we introduce an additional boundary symplectic potential that depends on $\delta X$.
The pullback of $X$ by $Y$ does not commute with the variation $\delta$.
Instead we find, defining $\delta_X := \delta X \circ X^{-1}$ as in  \eqref{deltaX} and using the chain rule, that under diffeomorphism $\delta_X$ transforms as
\begin{equation}
\delta_X \to \delta(Y^{-1} \circ X) \circ (Y^{-1} \circ X)^{-1} = Y^*( \delta_X - \delta_Y).
\end{equation}
Again this equation is directly analogous to the Yang-Mills case \eqref{deltagstarphi}.
Using $\delta_X$ we can construct a boundary symplectic potential $\Theta_S$ to cancel the action of the diffeomorphism on $\Theta_\Sigma$.
Let us define 
\begin{equation}
\Theta_S [g, X, \delta X] = 
\int_{S} \pi_g [\delta_X],
\end{equation}
where we recall that $S=X(s)$.
Under a diffeomorphism it will transform as 
\begin{equation}
\Theta_{Y^{-1}(S)}[Y^*g, Y^{-1} \circ X, \delta (Y^{-1} \circ X)] = \int_{S} \pi_g[\delta_X - \delta_Y].
\end{equation}
Addition of this boundary term to the symplectic structure will therefore cancel the unwanted term in the transformation of the bulk symplectic potential \eqref{grboundaryterm}.
We therefore have the diffeomorphism-invariant symplectic potential given by
\bea    
\Theta[g, \delta g, X, \delta X] &=& \Theta_{\Sigma} [g, \delta g] + \Theta_{S} [g, X, \delta X] \nonumber \\
&=& \int_{\Sigma} \theta[g, \delta g] + \oint_{S} \pi_g[\delta_X].
\eea
The $X$ dependence of this expression is implicit in $\Sigma=X(\sigma)$, $S= X(s)$, and in the definition of $\delta_X$.
This symplectic potential satisfies by construction the invariance
\be\label{diffinv}
\Theta[Y^*g, \delta Y^*g,  (Y^{-1} \circ X), \delta (Y^{-1} \circ X)]
=\Theta[g, \delta g,  X, \delta X].
\ee

We can express the boundary symplectic potential in a more familiar tensorial form using the binormal to the codimension-2 surface $S$.
Let $n^a$ be a timelike unit normal to $S$, and $s^a$ a spacelike unit normal.
We can then form the unit binormal $n_{ab} = (n_a s_b - n_b s_a)$, which is independent of the choice of normals $n$ and $s$.
The densitised unit binormal is $\epsilon_{ab} = \sqrt{q} \, n_{ab} \, \rd^{D-2}\sigma$ where $\sigma^\alpha$ are coordinates on $S$, and $q$ is the determinant of the induced metric $q_{\alpha \beta}$ on $S$.
The boundary symplectic potential $\Theta_S$ can be expressed as
\be
\Theta_{S} [g, X, \delta X] 
= \frac{1}{2} \oint_S \star \rd g \delta_X = \frac{1}{2} \oint_S \epsilon_{ab} \nabla^a \delta_X^b.
\ee

\subsection{Gauge invariance and Hamiltonian generators}
\label{subsection:gravity-puregauge}

We now want to explore the gauge invariance of the extended symplectic structure. 
This follows closely the formal derivation given in section \ref{formalgauge} for Yang-Mills.
Let $V$ be a vector field and consider the following variation
\be
\delta_V g_{ab} = {\cal{L}}_V g_{ab},\qquad 
\delta_V X^a =  - V^a.
\ee
This is the infinitesimal version of the transformation $(g,X)\to (Y^*g,Y^{-1}\circ X ) $ where $ Y = 1 + V + O(V^2)$. 
The infinitesimal version of diffeomorphism invariance of the symplectic potential shown in \eqref{diffinv} translates into the identity 
\be
L_V \Theta =0.
\ee 
where the equality is valid  off-shell.
This shows that the Hamiltonian is given by $-I_V(\Theta)$ since 
\be
I_V \Omega = L_V \Theta - \delta I_V \Theta = -\delta [I_V(\Theta)].
\ee
We are therefore left with the evaluation of
\be
I_V\Theta = \int_\Sigma \theta[g, {\cal L}_V g] - \int_S \pi_g[V]
= \int_\Sigma \epsilon_a R^a{}_b  V^b \hateq 0.
\ee
where we have used the identity \eqref{IT}. 
This shows that the generator of gauge transformation vanishes on shell even for gauge transformations with nontrivial support at the boundary.

\subsection{Bulk and boundary symplectic forms}

Having found $\Theta$ we can now evaluate the symplectic form $\Omega = \delta \Theta$, which determines the Poisson brackets between physical observables.
There will be two terms in this expression, one coming from $\Theta_\Sigma$, and the other from $\Theta_S$.
Here there is a new subtlety in the derivation that was not present in the Yang-Mills case.
Let us first recall that $\sigma$ is a surface in $\mathbb{R}^D$ with boundary $s$ and that we define $X(\sigma) = \Sigma$ and $X(s) = S$ as their images in spacetime.
The new subtlety is that the bulk symplectic potential is an integral over the surface $X(\sigma)$ and this leads to nontrivial variation of the bulk symplectic potential under variations of $X$.
This is in contrast with Yang-Mills where the bulk symplectic potential \eqref{YMtheta} does not depend on the choice of gauge.
This dependence on $X$ leads to $\delta_X \delta g$ terms in the variation of $\Theta_\Sigma$.
However we will see that these terms are all total derivatives, and hence can be expressed as boundary integrals.
Once this is done the symplectic form consists of a bulk integral containing $\delta g \delta g$ terms, and a boundary integral containing $\delta g \delta_X$ and $\delta_X \delta_X$ terms.
This structure is similar to that of Yang-Mills theory.

In order to vary an expression such as $\Theta_\Sigma$, which is an integral over the surface $\Sigma = X(\sigma)$, we first use the pullback by $X$ to express it as an integral over $\sigma$:
\begin{equation}
\int_{\Sigma} \theta = \int_\sigma X^*(\theta).
\end{equation}
Then we need only vary the integrand.
To vary the pullback of a tensor-valued variational form such as $\theta$ we make use of the identity \eqref{pullback}:
\begin{equation} \label{pullbackagain}
\delta X^*(T) 
= X^*(\delta T + \mathcal{L}_{\delta_X} T) 
\end{equation}
Recall that $\delta_X^a= \delta X^a\circ X^{-1}$ is a vector field associated with the variation $\delta X$.

Using \eqref{pullbackagain} to vary the bulk symplectic potential $\Theta_\Sigma$, we find
\begin{equation} \label{deltathetasigma}
\delta \Theta_\Sigma = \int_{\Sigma} \delta \theta + \int_{\Sigma} \mathcal{L}_{\delta_X} \theta.
\end{equation}
The first integral is the usual bulk symplectic form for gravity \cite{Crnkovic:1987tz}.
It is obtained as the metric variation of the symplectic potential $\Theta_\Sigma$ and therefore contains only $\delta g \delta g$ terms:
\begin{equation}
\Omega_\Sigma = \int_{\Sigma} \delta \theta. 
\end{equation}
Explicitly, it is given by (see \cite{Hollands2012}):
\begin{equation}
\Omega_\Sigma = \int_{\Sigma} \left(
\frac12 \delta g^{cd} \nabla^a \delta g_{cd}
+\frac12 \delta g^{ab} \nabla_b \delta g 
- \delta g^{cd} \nabla_d \delta g_c{}^a
+\frac12 \delta g \nabla_b \delta g^{ab}
-\frac12 \delta g \nabla^a \delta g
\right) \epsilon_a
\end{equation}
Here we use again the abbreviated notation used in \eqref{theta}.

The expression can then be evaluated in terms of the metric tangent to the slices and the normal accelerations \cite{Hawking:1998jf,Brown:2000dz, Freidel:2013jfa}. 
If we choose a slice $\Sigma$ defined by the condition $T= \text{constant}$ and equipped with a unit normal one-form $n_a \rd x^a = -N \rd T$, where $N$ is the lapse we can, following \cite{Freidel:2013jfa}, write the bulk component of the symplectic structure explicitly in terms of the induced metric $h$ and extrinsic curvature $K_n^{a b} := h^{ac} h^{bd} (\nabla_c n_d) $. 
If $\Sigma$ has no boundary it is given by 
\beq\label{OmegaBulk}
\Omega_\Sigma =
\int_\Sigma \delta \Pi^{ab} \delta h_{ab} + \delta \Pi_T \delta T,
\eeq
where the momenta conjugate to $(h_{ab}, T)$ are given by 
\beq\label{pi}
\Pi^{ab}:= \sqrt{h}\left(K_n^{ab}- K_{n} h^{ab}\right),\qquad 
\Pi_T := -\sqrt{h}\left(\Delta N \right),
\eeq
where $\Delta = D_a h^{ab} D_b$ is the spacelike Laplacian.
The second term in (\ref{OmegaBulk}) 
says that the Laplacian of the lapse is canonically conjugate to the time variable. This seems unusual to promote the lapse to a kinematical variable.
The reason it is usually not considered is because in the ADM approach, one fixes the foliation from the outset and focuses solely on hypersurface-normal deformations.
These are only a  subset of the deformations --- those that preserve the foliation --- hence they correspond to deformations for which $\delta T=0$.
Once we do this and ignore the canonical pair $(\Delta N , T)$ the lapse becomes simply a  Lagrange multiplier for the Hamiltonian constraint. 
Note however that the restriction to hypersurface-normal deformations is the reason why the ADM algebra is not equivalent to the algebra of spacetime diffeomorphisms, since in general a diffeomorphism changes the foliation.

The second term in \eqref{deltathetasigma} is a mixed term that depends on $\delta_X$ and $\delta g$. To analyse it we use Cartan's formula,
\begin{equation} \label{cartan}
\mathcal{L}_{\delta_X} \theta = i_{\delta_X} \rd \theta + \rd i_{\delta_X} \theta.
\end{equation}
The first term in \eqref{cartan} is proportional to $\rd \theta = \star \frac12(\nabla^a \nabla^b \delta g_{ab} - \nabla^2 \delta g)$ which vanishes for on-shell variations by the trace of the linearized vacuum Einstein equation.
It is therefore  a pure boundary term when $\delta g$ is an on-shell variation.
In general, when we do not restrict to on-shell variations,  we note that the term $i_{\delta_X} \rd \theta$ in the symplectic structure is an unfamiliar feature of the canonical formalism for gravity. Because $\delta_X$ appears only in the interior product, this term depends on the normal component of $\delta_X$ and hence arises from changes of the foliation. 
As such, it is similar to the $\delta T$ term discussed in the preceding paragraph and disappears if one restricts to hypersurface-normal deformations only.
Clearly, this issue of general deformations deserves further investigation, but for now, and because it doesn't play any role in our boundary study, we are content to consider variations for which this term vanishes.

We can now focus on the boundary terms in the symplectic form.
The second term  in \eqref{cartan} is manifestly a total derivative, and so can be written as an integral over $S$:
\begin{equation} \label{ldxtheta}
\int_{\Sigma} \rd i_{\delta_X} \theta = \int_{S} i_{\delta_{X}} \theta.
\end{equation}

We also have to vary the boundary symplectic form, which can be accomplished using \eqref{pullbackagain} again:
\begin{equation} \label{deltaThetaS}
\delta \Theta_S = \frac12 \int_{S} \delta (\star \rd g \delta_X) + \mathcal{L}_{\delta_X} (\star \rd g \delta_X).
\end{equation}
We can simplify this expression 
by expanding the middle term using Cartan's formula and discard the total derivative, since the surface $S$ has no boundary.
The result is
\begin{equation}
\delta \Theta_S = \frac12 \int_{S} \delta (\star \rd g) \delta_X + i_{\delta_X} \rd \star \rd g \delta_X + \star \rd g \delta(\delta_X).
\end{equation}

Combining this variation of $\Theta_S$ with the boundary integral from the variation of $\Theta_\Sigma$ \eqref{ldxtheta} we arrive at the boundary symplectic form:
\be \boxed{ \label{OmegaSpi}
\Omega_S = \frac12 \int_{S} 2 i_{\delta_X} \theta +  i_{\delta_X} \rd \star \rd g \delta_X  + \delta(\star \rd g) \delta_X +  \star \rd g \delta (\delta_X).
}
\ee
The first and third terms in this equation are the mixed $\delta g \delta_X$ terms, while the second and fourth  are the $\delta_X \delta_X$ terms.
The bulk piece contains only terms of the form $\delta g \delta g$ if we restrict either to on-shell or to hypersurface-normal deformations.
Thus we find a structure similar to that of Yang-Mills theory; the boundary terms in the symplectic structure lead to nontrivial Poisson brackets between $X$ and certain components of the metric in the neighbourhood of $S$.
Note that we can write this symplectic structure using the map $\pi_g[V] =\tfrac12 \star \rd g (V)$ as 
\be \boxed{ \label{OmegaS}
\Omega_S =  \int_{S}  i_{\delta_X} \left( \theta + \rd \pi_g[ \delta_X] \right)  + \delta(\pi_g) [\delta_X] -\tfrac12  \pi_g[[\delta_X,\delta_X]]
}
\ee
A significant difference between Yang-Mills and gravity is that the gravity symplectic potential depends on derivatives of $\delta X$ on the boundary: this shows that there will be a larger group of physical transformations that transform the metric in a first-order neighbourhood of the surface $S$.
Moreover we expect the Poisson brackets between observables to contain derivatives of the generators, unlike in Yang-Mills.

\subsection{Surface symmetries}

As in gauge theory, we can distinguish between two types of transformations.
The transformations considered in section \ref{subsection:gravity-puregauge} are diffeomorphisms, they simply relabel the points of $M$ and hence are pure gauge.
This is reflected in the fact that these transformations are null directions in the symplectic form.
However, there is a different class of transformations we can consider that transforms the reference frame $X^a$.
These transformations are genuine physical symmetries of the phase space and have nontrivial generators that we will derive in this section.
We will refer to them as surface symmetries, because they act at the entangling surface.

Before describing the surface symmetries, we first recall the action of gauge transformations.
A diffeomorphism acts on the metric by the pullback, and acts on $X$ on the left:
\begin{equation}
g_{ab} \to Y^*(g)_{ab}, \qquad X \to Y^*(X) = Y^{-1} \circ X.
\end{equation}
If we let $Y = I + V + O(V^2)$ where $V$ is a vector field on $M$, then $Y^{-1} \circ X = X - V \circ X + O(V^2)$. This means that infinitesimally we have  $I_V \delta X = -V \circ X$, and so $I_V \delta_X = -V$.
Thus we have the infinitesimal transformation 
\begin{equation}
I_V \delta g_{ab} = \mathcal{L}_V g_{ab}, \qquad I_V \delta_X = -V.
\end{equation}
Recall that $X$ is a map from an open set $U \subset \R^D$ into $M$, and a diffeomorphism $Y$ is a map $M \to M$ that relabels the points.
Because of the transformation rule for $X$, the point $X(x)$ still refers to the same physical point after a diffeomorphism.
This transformation is just a relabelling of points in $M$, and this explains why it is pure gauge.

Instead we will consider a transformation of the reference system $Z : U \to U$.
This corresponds to a change of the reference surface $s$ which is mapped onto $S$, and of the surroundings of that surface.
It acts by changing the labelling of the points, but keeps the dynamical fields unchanged:
\begin{equation}
g_{ab} \to g_{ab}, \qquad X \to X \circ Z.
\end{equation}
Note that like in Yang-Mills the symmetry acts on the opposite side of $X$ from the gauge transformation.

To find the infinitesimal version of this transformation, we let $Z = I + w+O(w^2)$, where $w$ is a vector field on $U$.
Then we have in components $(X \circ Z)^a = X^a + {\partial_b X^a} w^b +O(w^2)$. 
This defines a vector field $\Delta_w$ on phase space where 
\begin{equation} \label{IDeltaw}
I_{\Delta_w} \delta g_{ab} = 0, \qquad I_{\Delta_w} \delta_X^a =  {\partial_b X^a} ( w^b \circ X^{-1}) := W^a.
\end{equation}
Here we use upper case $W$ to denote the pushforward of the vector field $w$ to a vector field on $M$, i.e. $W = X_*(w)$, where $X_*: T U \to TM$ is the pushforward.
The infinitesimal transformation \eqref{IDeltaw} is the analog of the transformation generated by the electric field in Yang-Mills \eqref{Deltaalpha}.

Note that because $W$ depends implicitly on $X$, it has a nontrivial variation $\delta W$, which we will need below.
This follows from the identity $X^*(W) = w$, and our formula for the variation of the pullback. 
Varying this identity we find 
\begin{equation}
0 = \delta X^*(W) = X^*( \delta W + \mathcal{L}_{\delta_X} W )
\end{equation}
which implies
\begin{equation} \label{deltapushforward}
\delta W = - \mathcal{L}_{\delta_X} W = [W,\delta_X].
\end{equation}

\subsection{Classification of surface symmetries}

We now have a boundary symplectic structure for general relativity analogous to that of Yang-Mills theory. 
We have seen that as in Yang-Mills, the extension of the phase space via the introduction of boundary coordinates allows us to construct two types of canonical transformations: the gauge transformations whose Hamiltonian generators vanish on shell and the surface symmetries that act purely on the boundary variables. 
The latter are associated with a nonvanishing Hamiltonian generator that we will construct in the following section.
The key difference between Yang-Mills and general relativity is that the Yang-Mills symmetry generators are spacetime scalars, while the symmetry generators in general relativity are vectors.
This leads to a classification of different types of symmetries, depending on how the vector field $W$ meets the surface $S$.
To fix the idea it is convenient to introduce a local coordinate system $(x^i,\sigma^\mu)$ where $(x^i)_{i=0,1}$ denote coordinates normal to the surface and $(\sigma^\mu)_{\mu=1,\cdots, D-2}$ denote coordinates tangential to the surface which is located at $x^i= \text{constant}$.

Given an infinitesimal symmetry generated by a vector field $W$, we can divide its restriction to the surface $S$ into its normal part 
$W_\perp  ={W}^i_\perp\partial_{x^i}$ and tangential part 
$W_\parallel = W_\parallel^\mu \pa_{\sigma^\mu}$ as $W =  W_\parallel +  W_\perp$.
The important point is that  the contraction of the symplectic structure with the generator $\Delta_w$ depends not only on the value of $W$ on $S$ but  also on the normal derivatives of $W_\perp$, so that even when the normal component vanishes on $S$, it will have a nontrivial generator.
Vector fields whose parallel component vanish $W_\parallel\circeq 0$ and whose normal component and its first derivative in the normal direction also vanishes ${W}^{i}_\perp \circeq \partial_{j}{W}^{i}_\perp \circeq 0$, where $\circeq$ denotes equality on the surface $S$, are in the kernel of the symplectic structure and therefore pure gauge.
Thus we only implement as Hamiltonian transformations those vector fields that are nonvanishing in a first-order neighbourhood of $S$.

We can classify these surface symmetries into three classes:
\begin{itemize}
\item \textbf{Surface boosts} These are generated by vector fields with $ W_\parallel \circeq 0$ and $ W_\perp \circeq 0$ but $\partial_i W^j_\perp \not\circeq  0$. 
They generate position-dependent linear deformations of the normal plane of $S$. 
We call them surface boosts because they are a close analog of the boost transformations in the normal plane $(x^0,x^1)$.
Like the usual boost, which leaves invariant a codimension-2 plane, the surface boosts leave $S$ invariant.
Unlike the usual boost, the surface boosts are not isometries of any particular background.
Moreover we will see that these surface boosts generate a larger $\slr$ subalgebra of the two-dimensional linear group at each point of $S$.
\item \textbf{Surface diffeomorphisms} Vector fields with $ W_\perp \circeq 0$ and $W_\parallel \not\circeq 0$ generate infinitesimal diffeomorphisms that map the surface $S$ onto itself.
\item \textbf{Surface translations} The transformations with $ W_\perp  \not\circeq 0$ transform the surface $S$ normal to itself. 
\end{itemize}
Here we will focus on the class of surface-preserving transformations, i.e. the transformations such that $W_\perp \circeq 0$. 
These consist of the surface boosts and surface diffeomorphisms, which do not move the surface $S$. 
This is appropriate to the study of an entangling surface which is kept at a fixed location in spacetime, while we allow the angle at which the foliation $\Sigma$ meets $S$ to change.
Previous works have focused on the surface translations (which have been called ``would-be pure gauge'' degrees of freedom \cite{Carlip:1994gy}), but have largely overlooked the surface-preserving generators.
We leave a full treatment (treating all surface symmetries together) as an interesting area for future work, and comment on some of the technical challenges in the discussion.

\subsection{Surface symmetry generators}

We now consider the generators $H_w$ of the surface symmetries $\Delta_w$. 
These are obtained by solving the equation $\delta H_w = I_{\Delta_w} \Omega$. 
We first find $I_{\Delta_w} \Omega$, which has both a bulk component and a boundary component.
The bulk component $I_{\Delta_w} \Omega_\Sigma$ vanishes, because the transformation $\Delta_w$ vanishes when acting on the metric. 
We are left with the boundary term, which is:
\begin{align} \label{IDeltaWOmega}
I_{\Delta_w} \Omega &= I_{\Delta_w} \Omega_S \nonumber \\
&= I_{\Delta_w} \int_{S} \Big(  i_{\delta_X}\left( \theta  + \rd \pi_g[ \delta_X] \right) +  (\delta \pi_g)[\delta_X] -\tfrac12  \pi_g[ [\delta_X,\delta_X]] \Big) \nonumber \\
&=  \int_{S} \Big(  i_W \left( \theta  + \rd \pi_g[ \delta_X]\right) - i_{\delta_X} \rd \pi_g[W] -  (\delta \pi_g)[W] - \pi_g[ [W,\delta_X]] \Big)
\end{align}
Here we have made use of the identity $I_{\Delta_w} \delta (\delta_X) = -\frac12 I_{\Delta_w} [\delta_X, \delta_X] = [\delta_X, W]$.
Here $(\delta \pi_g)$ denotes the variation of the metric terms that appear in $\pi_g$ without variation of the argument.

We see that this interior product contains two types of terms, one associated with surface-preserving symmetries and one associated with translations. 
The first one is the translation contribution
\begin{equation}
 \int_{S}   i_W \Big(\theta + \rd \pi_g[\delta_X] \Big).
\end{equation}
This term vanishes for surface-preserving symmetries.  
Indeed for surface-preserving symmetries, $W$ is tangent to $S$, and hence the interior product $i_W \alpha$ vanishes when pulled back to $S$, for any form $\alpha$. 

From now on we specialize our analysis to the case of surface-preserving symmetries.
In this case the previous term vanishes and we are left only with the surface-preserving contribution
\begin{equation} \label{IdeltaWperp}
I_{\Delta_w} \Omega = - \int_S \Big(  (\delta \pi_g)[W] + \mathcal{L}_{\delta_X} \pi_g[W] + \pi_g[[W,\delta_X]] \Big) .
\end{equation}
where we have used the definition of $\pi_g$ and  discarded a total derivative term in order to write the second term as a  Lie derivative.
Using the identity \eqref{deltapushforward} for $\delta W$,
and the variational identity \eqref{pullbackagain}, we clearly see that this contraction is Hamiltonian that is $\delta H_w = I_{\Delta_w} \Omega$,
where 
\begin{equation} \label{Hw}
H_w =-\int_S \pi_g[W]= - \frac{1}{2} \int_S \star \rd g W.
\end{equation}
$H_w$ is therefore the Hamiltonian generator of the surface-preserving transformations.

We can now easily find the algebra of the surface-preserving transformations.
If $v$ is another surface-preserving transformation, then by direct substitution we have
\begin{equation} \label{HvHw}
\{ H_v, H_w \} = -I_{\Delta_v} \delta H_w = H_{[v,w]}.
\end{equation}
Note that the surface-preserving condition is preserved by the Lie bracket: if $v$ and $w$ are surface-preserving then so is $[v,w]$.
We also observe that the surface-preserving symmetries do not acquire a central charge at the classical level.
We will give a more explicit characterization of this algebra and its Poisson brackets in the following subsection.

We also note that our Hamiltonian \eqref{Hw} coincides with the Noether charge of Refs.~\cite{Wald1993,Iyer1994}.
However the context here is different: $W$ is not assumed to be an isometry of any fixed background, and $S$ need not be a section of a Killing horizon. 
Instead $W=X_*(w)$ parametrises an infinitesimal change of coordinate frame of the boundary surface.

\subsection{Algebra of surface symmetries} \label{section:twoplustwo}

 In order to describe explicitly the algebra of surface symmetries and its generators, we make a ``2+2'' decomposition\footnote{Of course, we are in general dimension $D$, so it is really a $2 + (D-2)$ decomposition.} of the metric in a neighbourhood of $S$, and evaluate the generator $H_w$ of a surface-preserving symmetry $w$.
In terms of the timelike unit normal $n$ and spacelike unit normal $s$ to $S$, the generator takes the form
\begin{align}
H_w &= -\frac12 \int_S \sqrt{q} (n^a s^b - n^b s^a) \nabla_a W_b \nonumber \\
&= -\frac12 \int_{S} \sqrt{q}\left( \nabla_{\n}({W}_\perp\!\cdot\!\s)- \nabla_{\s}({W}_\perp\!\cdot\!\n) -{W}^\mu_{\parallel } [\n,\s]^\nu q_{\mu \nu}\right) \nonumber \\
&= -\frac12 \int_{S} \sqrt{q}\left(\epsilon^{AB} \nabla_{\n_A}({W}_\perp\!\cdot\!\n_B) -{W}^\mu_{\parallel } q_{\mu \nu}  [{\n},\s]^\nu \right). \label{HV}
\end{align}
We have introduced the notation $\n_A =\n_A^a \pa_a$ for the two unit normal vectors $\n_0 = \n$ and $\n_1 = \s$, and the antisymmetric Levi-Civita symbol $\epsilon^{AB}$ with $\epsilon^{01} = 1$.
The vector $ [\n, \s]^\nu q_{\nu}{}^\mu$ is the projection of the Lie bracket of $n$ and $s$ onto the tangent plane of $S$ and it measures the integrability of the normal planes. 
Indeed the tangent bundle  at $S$ can be composed into the vectors tangent to $S$ and their normal $T_S M= TS \oplus (TS)^\perp$. By the Frobenius integrability condition, the normal sub-bundle $(TS)^\perp$ can be described as the tangent bundle of a normal submanifold of spacetime if and only if the Lie-bracket closes $[(TS)^\perp,(TS)^\perp]\subset (TS)^\perp$. In other words,  $[\n, \s]^\nu q_{\nu}{}^\mu=0$ is equivalent to the Frobenius integrability condition that the $(n,s)$ plane is tangent to a 2-dimensional submanifold of spacetime.

To evaluate the Hamiltonian  explicitly in terms of components of the metric we introduce coordinates $(x^i,\sigma^\mu)$ adapted to the $2+2$ decomposition.
Let $\sigma^\mu$ be coordinates on $S$, and let $(x^{i})_{i=0,1}$ be two coordinates in the normal directions.
In these coordinates we can parametrize the metric as
\be
\rd s^{2} = h_{ij}\rd x^{i}\rd x^{j} + q_{\mu \nu}\left(\rd \sigma^\mu - A^\mu_{i}\rd x^{i}\right)
\left(\rd \sigma^\nu - A^\nu_{j}\rd x^{j}\right).
\ee
Here $q_{\mu \nu}$ is the induced metric on $S$, $h_{ij}$ is a generalized lapse which defines the normal geometry and $A_i^\mu$ is a generalized shift.
We can view $A_i^\mu$ as a normal connection, since under relabelling $\delta \sigma^\mu =\varphi^\mu$ it transforms as a connection $ \delta A_i^\mu = \partial_i \varphi^\mu + [A_i,\varphi]^\mu$, where the bracket is the Lie bracket for vectors tangent to $S$.
The relationship between the normals $\n^A$ and the normal coordinates $x^i$ is given by a $2 \times 2$ matrix with coefficients $n^a_{i}$:
\be
\n^{A}= n^{A}_{i} \rd x^{i},\qquad h_{ij} = n^A_i\eta_{AB}n^B_j,
\ee
where $\eta_{AB}=\mathrm{diag}(-1,+1)$ is a flat 2D normal metric.
This expresses the fact that $\n^A$ is normal to the the level sets of constant $x^{i}$.
It is also necessary to express the normal vector fields $\n_A = \n_A^a \partial_a$ in terms of the metric coefficients:
\begin{equation}\label{normal}
\n_A = (n^{-1})^{i}_{A}(\partial_{i}+A_{i}^{\mu}\partial_{\mu}).
\end{equation}
Indeed we can readily check that $i_{\n_A} (\n^B) =\delta_A^B$ and that 
$i_{\n_A}(\rd \sigma^\mu - A^\mu_i\rd x^i) = 0$.

We can now further evaluate the Hamiltonian \eqref{HV} in terms of the 2+2 decomposition of the metric
\bea
\nabla_{\n_A}({W}_\perp\!\cdot\!\n_B) &=&(n^{-1})^{i}_{A}(\partial_{i}+A_{i}^\mu \partial_\mu)({W}^{j}_\perp n_{j B}) \nonumber \\
& \circeq &(n^{-1})^{i}_A n_{j B}(\partial_{i}+ A_{i}^\mu \partial_\mu)({W}^{j}_\perp ) \nonumber \\
& \circeq &h_{jk}(n^{-1})^{i}_A (n^{-1})^{k}_B
(\partial_{i} {W}^{j}_\perp  )
\eea
where we have used that $W_\perp \circeq 0$. 
Contracting with $\epsilon^{AB}$ we find
\begin{equation}
\epsilon^{AB} \nabla_{\n_A}({W}_\perp\!\cdot\!\n_B)
\circeq -\frac{1}{{\det(n)}}h_{jk}\epsilon^{ki} (\partial_{i} {W}^{j}_\perp  ).
\end{equation}
It will be convenient to introduce the densitised normal metric 
\be\label{H}
H_{j}{}^i :=
\frac{\sqrt{q}}{\det(n)} h_{jk}\epsilon^{ki}.
\ee
It is traceless by construction. 
Moreover, since $\det(h) = -(\det n)^2$, and $\det(\epsilon)=1$, we find that
\beq
\det H = -\det q,\qquad \Tr(H) =0.
\eeq
Using the form (\ref{normal}), we can also evaluate the twist density vector
\bea
\sqrt{q} [n,s]^\mu q_\mu{}^\nu  &=&  \sqrt{q} (n^{-1})^{i}_{0}(n^{-1})^{j}_{1} [(\partial_{i}+A_{i}^\mu \partial_\mu),((\partial_{j}+A_{j}^\nu \partial_\nu)]^\nu \\
&=& \frac{\sqrt{q}}{\det({n})}
\left(\partial_{0}A_1^\nu - \partial_1 A_0^\nu +[A_{0}, A_1]^\nu \right) =: F^\nu.\label{F}
\eea
It is proportional to  the curvature of the normal connection.
The proportionality coefficient ${\sqrt{q}}/{\det({n})}$ is the same we used to rescale the normal metric.
From \eqref{H},\eqref{F} we see that the Hamiltonian can be written in terms of the densitised normal metric and curvature as 
\be
H_{w} =\frac12\int_{S} 
\left( H_i{}^j
(\partial_{j}{W}^{i}_\perp  )
 +{W}^{\mu}_\parallel F_{\mu}\right) .
\ee

We can now see that the normal component of the Lie bracket for surface-preserving transformations becomes the $\mathfrak{sl}(2,\mathbb{R})$ algebra while the tangential component reduces to the $(D-2)$-dimensional diffeomorphism group acting on $S$:
\bea
\partial_{i}[V,W]^{j} &\circeq& \partial_{i}{V}^{k}_\perp\partial_{k}{W}^{j}_\perp -\partial_{i}{W}^{k}_\perp \partial_{k} {V}^{j}_\perp
+  V^\mu_\parallel \partial_\mu (\partial_i  W^j_\perp) 
-  W^\mu_\parallel \partial_\mu (\partial_i  V^j_\perp)
\\
{[V,W]}^\mu &\circeq&
{V}^\nu_\parallel\partial_\nu{W}^\mu_\parallel- 
{W}^\nu_\parallel\partial_\nu {V}^\mu_\parallel .
\eea
from which we find the Poisson brackets between generators:
\bea
\tfrac12 \{H_i{}^j(\si),H_k{}^l(\si')\}&=& 
(\delta_i ^l H_k{}^j - H_i{}^l\delta_k^j)(\si)\delta^{(D-2)}(\si,\si'),\\
\tfrac12 \{F_\mu(\si),F_\nu(\si')\}&=& F_\mu(\si') \partial_\nu\delta^{(D-2)}(\si,\si')-
F_\nu(\si) \partial_\mu'\delta^{(D-2)}(\si,\si'), \\
\tfrac12 \{ H_{i}{}^j(\si), F_\mu(\si')\} &=& H_{i}{}^j(\si') \partial_\mu \delta^{(D-2)}(\si,\si').
\eea
We see that the components of the densitised curvature $F$ act as generators of tangential diffeomorphisms. 
They have a nontrivial action on $H_{i}{}^j$, which transforms as a scalar under diffeomorphism, while $H_{i}{}^j$ itself generates a local $\mathfrak{sl}(2,\mathbb{R})$ algebra.
Thus we see that the group of surface-preserving symmetries is $\text{Diff}(S) \ltimes \slr^S$.

Let us introduce $\mathfrak{sl}(2,\R)$ generators:
\begin{equation}
K_0 = \tfrac{1}{2} (H_{0}{}^1 - H_{1}{}^{0}), \qquad K_1 = H_{0}{}^0=-H_1{}^1, \qquad K_2 = \tfrac{1}{2}(H_{0}{}^1 + H_{1}{}^0).
\end{equation}
$K_0$ is an elliptic generator while $K_1$ and $K_2$ are hyperbolic.
They satisfy the $\mathfrak{sl}(2,\R)$ commutation relations
\begin{equation}
\{ K_0, K_1 \} =  2K_2, \qquad \{ K_1, K_2 \} = - 2K_0, \qquad \{ K_2, K_0 \} = 2K_1. 
\end{equation}
The determinant of $H$ is 
\begin{equation}
\det H = H_{0}^0 H_{1}^1 - H_{0}^1H_1{}^0 = K_0^2 - K_1^2 - K_2^2
\end{equation}
which is precisely the Casimir of $\mathfrak{sl}(2,\mathbb{R})$.
The unitary representations of $\slr$ come in a discrete series for which the Casimir is positive and continuous series for which it is negative.
Since we require $\det H$ to be negative, we are interested in the continuous series. Note that the previous relation means that the Casimir of $\slr$ is the area element
 \beq \label{casimir}
 \sqrt{q} = K_1^2+ K_2^2 - K_0^2.
 \eeq

Since the local $\slr$ symmetry is a new feature, we briefly comment on the relation between this symmetry and the more familiar boost symmetry of Carlip and Teitelboim \cite{Carlip1993}.
In that work it was argued that the near horizon symmetry consists of a single boost, and that the quantity canonically conjugate to the boost is the horizon area density $\sqrt{q}$.
The difference between the two symmetry groups comes because our $\slr$ transformations are not isometries of the normal geometry.
The relation between the two symmetries can be understood as a particular gauge-fixing.
The normal metric $h_{ij}$ transforms in the adjoint representation of $\slr$, and since the metric must be nondegenerate it will spontaneously break this symmetry.
It is natural to fix the metric to the flat metric $h_{ij} = \eta_{ij}$, in which case one should consider the little group of transformations that preserves $h_{ij}$. 
This gauge corresponds to the conditions $H_0{}^0=0$ and $H_0{}^1= H_1{}^0=-1$.
The little group has a single unbroken generator $K_2$ which generates boosts.
In this gauge, the generator becomes
\begin{equation}
K_2 = \frac12 (H_{0}{}^1 - H_{1}{}^0) = \sqrt{q},
\end{equation}
the area element of the entangling surface. 
Thus we recover Carlip and Teitelboim's result as a gauge fixing of ours.

The situation can be summarized with a simple analogy to the rotation group $\SO(3)$. 
The Casimir $J^2$ is of course different from the rotation generator $J_z$.
If we fix the axis of rotation then classically $J_z$ and $\sqrt{J^2}$ coincide, but from the point of view of the representation theory they are of course distinct and we should not conflate the two.
In particular, they have different spectra at the quantum level and the representation theory is controlled by the Casimir $J^2$.

\section{Quantization and entanglement}
\label{section:quantization}

So far we have constructed the algebra of boundary symmetries in Yang-Mills theory and in general relativity.
At the quantum level, we expect this algebra to be realized in the usual way as an algebra of operators on a Hilbert space.
In Yang-Mills theory, this is simply the algebra of surface charges on the entangling surface. 
In a lattice realization it is the 'electric center' algebra of Ref.~\cite{Casini:2013rba}, and it is represented on a Hilbert space spanned by generalized spin network states used in Ref.~\cite{Donnelly2011}.

In gravity, we do not have an explicit construction of the Hilbert space as we do in Yang-Mills.
In constructing such a Hilbert space it is extremely useful to have a group of symmetries as a guide.
Here we have constructed a large group of surface symmetries whose representation theory can guide the construction of the Hilbert space of a local region.
We therefore give a brief sketch of how to construct representations of the algebra of surface symmetries.
This appears to be a rather challenging problem whose full solution we are forced to leave to future work.

\subsection{Quantization}

Having found the algebra satisfied by the surface symmetries in gravity we can look briefly at its quantization, which amounts to studying representations of this algebra.
Our surface preserving symmetry group
$\text{Diff}(S) \ltimes \slr^S$ has the structure of a semidirect product, whose projective representations have been extensively studied by Mackey \cite{Mackey1989}.
This approach is a generalization of the familiar classification of projective representations of the Poincar\'e group $\SO(1,3) \ltimes \R^4$.
First one starts with irreducible representations of the translation group $\R^4$, which are labelled by momenta, and then defines the little group as the subgroup of the Lorentz group that preserves the given momentum vector.

In our case we have diffeomorphisms playing the role of the Lorentz group and $\slr^S$ playing the role of translations.
We first fix an irreducible representation of $\slr^S$, which amounts to choosing the value of the Casimir operator at each point of $S$. 
We will assume that this choice is made smoothly which means that by the relation \eqref{casimir} we specify an area form on $S$.
This area form is analogous to the choice of momentum, and the total area (which is diffeomorphism-invariant) is analogous to the mass.
Next we consider irreducible representations of the little group, which is the subgroup of diffeomorphisms that preserve the given area form, which we denote $\text{APD}(S)$, where APD stands for Area Preserving Diffeomorphism.
The choice of APD representation is analogous to the spin in the classification of Poincar\'e group representations.
Thus the irreducible representations of our algebra are labelled by a total area, and an irreducible representation of $\text{APD}(S)$.
Using this information, one can then construct the irreducible representations of the surface symmetry algebra by the method of induced representations.

The algebra of area-preserving diffeomorphisms $\text{APD}(S)$ of a surface $S$ has been studied particularly in the case where $S = S^2$ (which fortunately is also the most physically relevant case, as it would apply to both black hole and cosmological horizons in a four-dimensional spacetime).
In particular its structure constants are known \cite{deWit:1988ig}, and it is known that for a surface of genus $g$ there are $2g$ central charges.
It would therefore be of great interest to understand more completely the representations of $\text{APD}(S^2)$.

\subsection{Entangling product}

Provided we can construct Hilbert spaces associated to local regions, the next question is how to glue together two such Hilbert spaces along a common boundary to make the Hilbert space of a larger region.
The diffeomorphism-invariance of general relativity would appear to be an obstacle, since diffeomorphisms naively can move the location at which the two Hilbert spaces are glued.
Here we point out that such a gluing construction has effectively been carried out for a large class of rigorously-constructed diffeomorphism-invariant quantum field theories: topological quantum field theories (TQFTs).
Although these theories have no local degrees of freedom, they provide useful models of how we can glue subsystems together in a diffeomorphism-invariant theory.

In the axiomatic approach to TQFT, every closed codimension-1 surface $\Sigma$ is assigned a Hilbert space $Z(\Sigma)$, and each codimension-0 surface $M$ with boundary $\partial M$ is associated with a state in the Hilbert space $Z(\partial M)$ \cite{Atiyah1988}.
These satisfy a set of identities that allow one to calculate the partition function of a closed manifold $M$ by cutting along codimension-1 surfaces and gluing them back together.
This structure is sufficient to classify two-dimensional TQFTs \cite{SchommerPries:2011np}, but in higher dimensions this is not sufficient. 
This is essentially because codimension-1 surfaces are more complicated objects as the spacetime dimension increases.

One of the axioms of TQFT states that a disjoint union of two surfaces is assigned a tensor product of Hilbert spaces, $Z(\Sigma_1 \sqcup \Sigma_2) = Z(\Sigma_1) \otimes Z(\Sigma_2)$.
One can augment this structure by allowing codimension-1 surfaces to be cut along codimension-2 surfaces, the resulting structure is called an extended TQFT \cite{Baez1995}.
This additional structure is used to classify higher-dimensional extended quantum field theories in terms of $n$-categories \cite{Lurie2009}.

In general, the structure associated to a codimension-2 surface is a category, and a codimension-1 surface with boundary is a morphism in its boundary category.
However one can understand this better in a simple example, such as two-dimensional Yang-Mills theory \cite{Oeckl2006}.
Two-dimensional Yang-Mills is not quite a TQFT according to the usual axioms. First, its Hilbert space is infinite-dimensional, and second it is not invariant under the full diffeomorphism group but only under the subgroup of area-preserving diffeomorphisms.
However it is close enough to a topological quantum field theory that much of the axiomatic framework carries over to this case \cite{Cordes1994}.

In 2D Yang-Mills, the Hilbert space associated to a closed codimension-1 manifold (a circle) is the space of square-integrable class functions of the gauge group $G$.
These are functions $\psi(u)$ of the holonomy $u = \exp(i \oint A) \in G$ that are invariant under the gauge transformation $u \to g^{-1} u g$, i.e. $\psi(u) = \psi(g^{-1} u g)$.
One can also construct the Hilbert space associated to an interval, which is the space of square-integrable functions of the connection integrated along the interval $u = \exp(i \int A) \in G$.
In the case of the interval, the states are not restricted to be class functions.
The Hilbert space of an interval then carries extra structure: it carries two representations of $G$ corresponding to left and right multiplication of the group element $u$ acting as $u \to g u$ and $u \to u g^{-1}$ respectively.
This endows the Hilbert space of an interval with the structure of a module over the group algebra of $G$.

The Hilbert spaces of intervals are not glued together with the usual tensor product of Hilbert spaces. 
Recall that one defines the tensor product over $\mathbb{C}$ as the cartesian product $\mathcal{H}_1 \times \mathcal{H}_2$ modulo the equivalence $(\alpha v, w )\simeq( v,\alpha w)$ where $\alpha\in \mathbb{C}$.
In the case of a tensor product over an algebra ${\cal A}$, we define $\mathcal{H}_1 \otimes_{\cal A} \mathcal{H}_2$ such that $(v a) \otimes w = v \otimes (a w)$ for all elements $a \in {\cal A}$. When the algebra is the group algebra ${\cal A}=\mathbb{C}(G)$ we denote this entangling product 
$\mathcal{H}_1 \otimes_{G} \mathcal{H}_2$. 
As we have seen this entangling group and the corresponding product naturally appear in Yang-Mills and gravity as the surface symmetry group. 
Interestingly it has been conjectured in \cite{Witten2009} that the slice of a spacetime outside a black hole horizon should have precisely this structure. 
We see here that it appears in fact generically for an entangling surface in a gauge theory and in gravity.

\section{Discussion} \label{section:discussion}

We have defined a formalism for associating a classical phase space to a bounded region that applies to both Yang-Mills theory and general relativity.
We have restricted our analysis to {\it entangling surfaces}: codimension-2  surfaces which are fixed in spacetime.
By including the choice of gauge as a canonical coordinate, we obtain an extended phase space that is invariant under gauge transformations and which carries a representation of an induced boundary symmetry group $G_S$.
In Yang-Mills theory, the boundary degrees of freedom consist of a section of a principal $G$-bundle and the normal component of the electric field, which act like surface charges on the boundary.
In gravity, the boundary degrees of freedom consist of the location of an embedding surface $s \to X^a(s)$ and its normal and tangential geometry, both intrinsic and extrinsic.
The remarkable feature of such a construction is that it selects a subset of the boundary observables to be the charges for this emergent boundary symmetry. 
This is the normal electric field in Yang-Mills, and the normal conformal metric, normal curvature and area density in gravity.
While these components of the fields are commuting variables from the point of view of the bulk theory they become noncommutative in the presence of a boundary, forming a Lie subalgebra.
We have seen that the appearance of this boundary symmetry group allow us to define at the quantum level the total Hilbert space associated with the gluing of two region as the entangling product. 
The entangling product is a generalisation of the tensor product where states are restricted to be singlets under the boundary symmetry.
We have described a classical version of this entangling product, which takes the form of a symplectic reduction on the product of two phase spaces.

It is well-known in the case of lattice Yang-Mills that the identification of boundary charges as the normal electric field can be understood as a consequence of the discretisation. 
In a discretisation two facts are clear.
First, the discrete electric field generates the action of a nonabelian group and hence must be noncommutative. 
Second, one must demand continuity of the electric field, but not of the magnetic field across the discrete surface.
This is because the electric and magnetic fields are non-commuting variables and the identification $E_L=E_R$  generates a canonical transformation acting on the magnetic fields $B_L$ and $B_R$ under which admissible magnetic field observables must be invariant. 
This means that the discretisation requires a splitting of the boundary data into two halves.
One half becomes the nonabelian charges and the other half corresponds to the bulk conjugate variables. 
We can now appreciate why putting gravity on a lattice in a way that respects gauge symmetry is so challenging: it requires an understanding of how to split in half the metric variables along the discrete lattice, and it also requires us to understand what commutation relations these discrete variables should satisfy.
What is remarkable in our analysis is that we recover this splitting in two halves of the boundary variables and their commutation relations from an analysis of gauge invariance.
This suggest that our analysis should provide new clues into the problem of discretising gravity in a covariant manner. 
If one views the process of discretisation as a necessary step in order to define a quantum field theory nonperturbatively, this could be relevant for the construction of a theory of quantum gravity. 

In fact this reflexion leads to an interesting conjecture: in the continuum we have seen that preserving gauge symmetry in the presence of boundaries implies the existence of a boundary symmetry on entangling surfaces.
One can view the vertices of a discrete spacetime as a set of regions bounded by entangling surfaces, which are glued together along links of the lattice.
If we take this point of view we can ask what are the necessary conditions on the discrete variables to ensure that the resulting theory manifests diffeomorphism symmetry in the continuum limit?
A natural conjecture in view of our analysis is that each vertex of the discretisation needs to carry a representation of the entangling group $G_S$.
In other words, not only does the boundary symmetry appear as a result of gauge invariance, but gauge invariance could be seen to emerge as the result of preserving the local surface symmetry. 
This is a new and very tantalising possibility that deserves a proper analysis.

Our analysis so far has been restricted to entangling surfaces at a fixed position in the bulk of the spacetime. 
We have therefore restricted our attention to surface symmetries that preserve the location of the entangling surface.
But the local symmetry group also includes surface translations that move the entangling surface.
These transformations introduce new complications; in particular we have not yet found a Hamiltonian generator of surface translations analogous to that for surface-preserving transformations \eqref{Hw}.
We have laid the groundwork for such an analysis, but it needs to be completed. 
Previous work suggests that one will have to impose boundary conditions that fix the location of the surface in order to have a nondegenerate symplectic structure, as found in \cite{Carlip1999}.
It remains to be seen whether such an invariant specification of the boundary region is possible in general and whether it is  necessary and sufficient to complete the symplectic structure.

It would be very interesting to generalise our results to the case of more general surfaces  and to understand how our results connects to the results obtained when the boundary is at infinity.
Much recent work has focused on asymptotic symmetries of Yang-Mills theories \cite{Strominger2013a,He2014b} and gravity \cite{Strominger2013b,He:2014laa}.
In particular, \cite{Mohd:2014oja} showed how the asymptotic symmetries of electrodynamics may arise by augmenting the phase space with edge modes living at the ``edge of infinity''.
However, so far there is no construction of an asymptotically flat phase space for gravity that contains a representation of the BMS group, except in 2+1 dimensions \cite{Andrade:2015fna}.
The methods developed in this paper may be useful to construct a phase space that includes the soft photon and graviton modes on which the action of the asymptotic symmetries is Hamiltonian.
We have seen so far that our method constructs a phase space containing a representation of a large nontrivial subgroup of the diffeomorphism group.
It would therefore be interesting to impose asymptotically flat boundary conditions within our formalism, and to see what asymptotic symmetry group is obtained.

Another related question concerns the relationship of our analysis with the usual Israel junction conditions.
Our analysis shows that we need to identify the normal geometry across the codimension-2 entangling surface but not the tangential geometry (except for the area density), while the Israel junction condition (in the absence of a surface stress tensor) demands the continuity of the tangential metric and extrinsic curvature across a codimension-1 boundary.
While there is no immediate contradiction between these results since they apply to radically different boundary conditions (codimension-2 versus codimension-1 boundaries) they create an interesting tension that has to resolve itself when we consider dynamical boundaries.
In order to understand this we first need to complete the canonical analysis of surface translations.

In particular, this matching raises the following issue.
In electrodynamics, the division of the Hilbert space requires splitting electric field lines into positive and negative surface charges, which must cancel in pairs to create gauge-invariant states.
Does this mean that the corresponding matching condition in general relativity requires the introduction of both positive and negative energy\footnote{We thank Steve Giddings for emphasizing this point.}?
Here we can provide a resolution of this tension.
Consider for example a black hole spacetime with a bifurcation surface divided into two regions along this surface.
Then the energy is related to the generator of the Killing flow, which we have found is simply a component of the normal metric.
These metric components are matched up to a sign coming from the convention of choosing $n$ to be an outward-facing normal vector.
Thus there is no contradiction because the generators of surface boosts do not have any fixed sign.
On the other hand, the area element does need to be positive in order for the surface $S$ to be spacelike.
However we have identified this as the Casimir of $\slr$, and so the matching condition imposes that it must be equal on both sides.
Thus we see that our matching condition gives the expected positivity properties, and there does not seem to be any problem of negative energy.
We expect to find no further contradiction when considering the generators of surface translation, but this remains to be explored.

Here we have focused on Yang-Mills theory and general relativity, but our formalism appears to be flexible enough to handle a very general class of field theories.
It would be interesting to apply our formalism to more general theories of gravity, for example with higher curvature couplings.
As a simplest example, inclusion of a nonzero cosmological constant would allow us to make contact with gravity in asymptotically AdS space.
The algebra we derive is reminiscent of the commutation relations of a surface stress tensor, which is an important part of the AdS/CFT correspondence, though we have so far not imposed any boundary conditions.
Another interesting case for our formalism is Chern-Simons theory, where the edge states and their relation to entanglement entropy have been understood for some time.
There we anticipate nontrivial Poisson brackets between $\varphi$, which do not appear in pure Yang-Mills theory.
This may help to explain the appearance of quantum group symmetry, and may shed new light on the open problem of discretizing Chern-Simons theory, which has so far only been solved for particular lattices and gauge groups \cite{Turaev1992,Sun:2015hla}.

More generally, our work seems to be pointing toward a certain extension of the formalism of quantum field theory.
In topological quantum field theory (TQFT) there is a useful notion of an extended TQFT \cite{Baez1995}.
In extended TQFT, one associates Hilbert spaces not only to closed codimension-1 surfaces, but also to codimension-1 surfaces with boundary.
The codimension-2 boundaries of these surfaces are then given some additional structure that encodes how the Hilbert spaces of two regions with boundary may be glued together to make a larger region.
This additional structure seems to be precisely what is needed in order to define entanglement entropy, which suggests the utility of considering extended field theory beyond the TQFT context.
In addition to formalizing the notion of extended QFT, it also raises the interesting question of whether a quantum field theory can always be extended in this way and whether such an extension is unique.
Theories with electric-magnetic duality may provide an interesting example, as they can be viewed as gauge theories in two different ways, suggesting multiple extensions and hence multiple ways of defining the entropy for such theories.
The existence of different ways to define entanglement entropy in the algebraic definition of entropy were noted in \cite{Casini:2013rba}, and there may be an analogous freedom in the extended Hilbert space definition.
Another interesting case are certain quantum Hall states admitting distinct edge phases \cite{Cano:2013ooa}, further suggesting different ways of extending the same TQFT.

Ultimately, we would like to apply these classical considerations to the problem of entanglement entropy in quantum gravity.
In order to do this, we need a Hilbert space which is the quantization of our classical phase space.
This is a hard problem, but it may be easier than the usual quantization problem because we have a large physical symmetry group.
Thus an important step in the quantization is to find the irreducible representations of $\text{Diff}(S) \ltimes \slr^S$.
These irreducible representations play an important role in the entanglement entropy, as we will review below.
In particular, the irreducible representations label superselection sectors of the Hilbert space $\mathcal{H}_\Sigma$.
Note that the entanglement entropy for compact groups contains a term that depends on the dimension of the irreducible representations of the gauge group.
Since $\slr$ is noncompact, all irreducible representations are infinite-dimensional; presumably this leads to a divergence that must be regulated.
However the regularization of this divergence via heat kernel methods has been understood in other models with $\slr$ symmetry \cite{Gupta:2014hxa}, so we do not expect this to pose any great difficulty.

Another important question that affects the quantization is the existence of central charges.
These central charges appear at the classical level in AdS${}_3$ gravity \cite{Brown1986}, and underlie the derivation of BTZ black hole entropy \cite{Strominger:1997eq} and its generalization to higher dimensions \cite{Carlip1998}.
The derivation of \eqref{HvHw} shows that there is no classical central term in the algebra of surface-preserving transformations.
However the central charge in \cite{Carlip1998} comes from diffeomorphisms that move the surface, so we may expect central terms to appear once we include the surface translations.

Finally, our work can be seen as a first step towards the problem of defining entanglement entropy in nonperturbative quantum gravity.
Once we have identified the Hilbert spaces $\mathcal{H}_\Sigma$ carrying an action of the group $G_S$ of surface symmetries, we can define the entanglement entropy by the embedding
\begin{equation}
\mathcal{H} = \mathcal{H}_\Sigma \otimes_{G_S} \mathcal{H}_{\overline \Sigma} \subset \mathcal{H}_\Sigma \otimes_{\mathbb{C}} \mathcal{H}_{\overline \Sigma}.
\end{equation}
We can now define a density matrix $\rho_\Sigma$ by embedding states of $\mathcal{H}$ into $\mathcal{H}_\Sigma \otimes \mathcal{H}_{\overline \Sigma}$, and tracing over $\mathcal{H}_{\overline \Sigma}$.
Any density matrix $\rho_\Sigma$ defined in this way commutes with the action of $G_S$.
This leads to a large symmetry of the problem, and a significant simplification of the entanglement entropy.
To make use of this symmetry, we first decompose the Hilbert space into irreducible representations of $G_S$.
Letting $R$ denote the irreducible representations, we have $\mathcal{H}_\Sigma = \bigoplus_R \mathcal{H}_{\Sigma, R}$.
Since the reduced density matrix commutes with the action of the surface symmetries, we can also decompose it as $\rho_\Sigma = \bigoplus_R p(R) \rho_{\Sigma,R}$. 
Here $p(R)$ is the probability that the state lies in the superselection sector described by the representation $R$; this factor is needed to keep the normalization condition $\Tr(\rho_\Sigma) =1$ and $\Tr(\rho_{\Sigma,R}) = 1$.
The entanglement entropy in gauge theory then takes the form \cite{Donnelly2011}:
\begin{equation} \label{Sent}
S_\text{ent} = \sum_{R} p(R) \left(- \log p(R) + \log \dim R + S(\rho_{\Sigma,R}) \right).
\end{equation}
Thus the representation theory of the surface symmetry group $G_S$ and the identification of its generators already tells us a lot about the entanglement entropy.

It is natural to compare formula \eqref{Sent} for the entanglement entropy with the generalized entropy
\begin{equation} \label{Sgen}
S_\text{gen} = \frac{A}{4 G} + S_\text{out}.
\end{equation}
In both of these equations the entropy is a sum of an entropy associated with degrees of freedom localized away from the entangling surface (the term $S_\text{out}$ of the generalized entropy and the term $S(\rho_{\Sigma,R})$ of the entanglement entropy) and the expectation value of an operator (the Bekenstein-Hawking entropy term in the generalized entropy, and the term $\log \dim R$ in the entanglement entropy).
The quantity $\log \dim R$ is a gauge-invariant observable that depends on the fields at the entangling surface - for example, in Yang-Mills we can determine it explicitly as a functional of the normal electric field.
This suggests the possibility that the Bekenstein-Hawking entropy arises as the part of the entanglement entropy that is determined by the representation theory of the diffeomorphism group.
In other words, the Bekenstein-Hawking entropy of a black hole could arise from counting the different ways in which the interior can be glued to its exterior; our boundary symplectic form gives the appropriate measure for this counting.
There is of course much to be done to establish or refute this conjecture.
In particular, we have to study the representation theory of the surface symmetry group to establish the relationship between $\dim R$ and the geometry of the surface $\Sigma$.
While the appearance of the area density as the Casimir of local $\slr$ symmetry is a suggestive, the real test of quantum gravity is to obtain the area law with the correct finite prefactor of $1/4G$.

\section*{Acknowledgments}

We thank Steve Giddings, Dan Harlow, Ted Jacobson, Don Marolf, Djordje Minic, Rafael Sorkin, and Aron Wall for their comments and insight.
We also thank Lin-Qing Chen and Henrique Gomes for finding some typos, and Antony Speranza for finding an error in a previous version.
This research was supported in part by Perimeter Institute for Theoretical Physics. 
Research at Perimeter Institute is supported by the Government of Canada through Industry Canada and by the Province of Ontario through the Ministry of Research and Innovation.
WD is supported by the Department of Energy under Contract DE- SC0011702

\bibliographystyle{utphys}
\bibliography{symplectic}

\appendix

\section{Field-space exterior calculus} \label{appendix:exterior}

The purpose of this appendix is to gather some technical results about the calculus of exterior $(p,q)$ forms.

In the algebra of spacetime forms we have the exterior derivative $\rd$ and interior product $i_V$ from which we define the Lie derivative
\begin{equation}
\mathcal{L}_V = i_V \circ \rd + \rd \circ i_V
\end{equation}
In the algebra of field space forms we have an analogous exterior (variational) derivative $\delta$ and interior product $I$, which takes as argument an infinitesimal transformation of the fields.
In particular we will be interested in the case where its argument is a spacetime Lie derivative, so that $I_V \delta \phi = \mathcal{L}_V \phi$ where $\phi$ stands for any of the dynamical fields.
In particular, we define:
\begin{equation}
I_V \delta g_{ab} = \nabla_a V_b + \nabla_b V_a, \qquad I_V \delta_X^a = -V^a.
\end{equation}

The field-space interior product with a vector field $V$ satisfies the usual anti-Leibniz rule: when $\alpha$ is a field space $p$-form and $\beta$ is a field space $q$-form, we have
\begin{equation}
I_V (\alpha \beta) = (I_V \alpha) \beta + (-1)^p \alpha (I_V \beta).
\end{equation}
Note, however, that the interior product whose argument is a field-space one-form such as $\delta_X$  obeys the Liebniz rule
\begin{equation}
I_{\delta_X} (\alpha \beta) = (I_{\delta_X} \alpha) \beta + \alpha (I_{\delta_X} \beta).
\end{equation}
This rule follows from the antisymmetry property $\alpha \beta = (-1)^{pq} \beta \alpha$ and the fact that $I_{\delta_X}$ preserves the degree of forms (whereas $I_V$ decreases it).

One can define also the Lie derivative for field-space forms,
\begin{equation} \label{straightL}
L_V = I_V \delta + \delta I_V.
\end{equation}
This satisfies again the usual Liebniz rule
\begin{equation}
L_V(\alpha \beta) = (L_V \alpha) \beta + \alpha L_V(\beta).
\end{equation}
For the Lie derivative along $\delta_X$, we must define the Lie derivative with the opposite sign,
\begin{equation}
L_{\delta_X} = I_{\delta_X} \delta - \delta I_{\delta_X}.
\end{equation}
This leads to the graded Liebniz rule 
\begin{equation}
L_{\delta_X}(\alpha \beta) = (L_{\delta_X} \alpha) \beta + (-1)^p \alpha (L_{\delta_X} \beta)
\end{equation}
which reflects the fact that $L_{\delta_X}$ increases the form degree by one.

We now show that the field-space Lie derivative is related to the spacetime Lie derivative as:
\begin{equation} \label{straightLidentity}
L_V = \mathcal{L}_V + I_{\delta V}.
\end{equation}
We first show that it holds for scalars and the coordinate one-forms $\delta \phi$, and then extend this to arbitrary products.
For a field-space $0$-form $f$ we have
\begin{equation}
L_V(f) = I_V \delta f = \mathcal{L}_V f = (\mathcal{L}_V + I_{\delta V}) f
\end{equation}
and a coordinate one-form, we have
\begin{equation}
L_V(\delta \phi) = \delta I_V \delta \phi = \delta (\mathcal{L}_V \phi) = \mathcal{L_V} \delta \phi + \mathcal{L}_{\delta V} \phi = (\mathcal{L}_V + I_{\delta V}) \delta \phi.
\end{equation}
Now suppose that forms $\alpha$ and $\beta$ both obey the identity \eqref{straightLidentity}. 
Then their product does also, which follows from the fact that both $L_V$ and $\mathcal{L}_V + I_{\delta V}$ satisfy the Liebniz rule:
\begin{align}
L_V(\alpha \beta) &= L_V(\alpha) \beta + \alpha L_V(\beta) \nonumber \\
&= (\mathcal{L}_V \alpha + I_{\delta_V} \alpha) \beta + \alpha (\mathcal{L}_V \beta + I_{\delta_V} \beta) \nonumber \\
&= (\mathcal{L}_V + I_{\delta V})(\alpha \beta).
\end{align}
This completes the proof of \eqref{straightLidentity}.

For a transformation such as $L_{\delta_X}$, the identity \eqref{straightLidentity} is slightly modified.
In this case $L_{\delta_X}$ is an infinitesimal transformation of the fields that depends linearly on $\delta_X$, so it is a vector-valued differential form. 
The Lie derivative can be naturally extended to vector-valued forms (see \cite{Kolar1999}) and we can carry out the same generalization to phase space.
Since $\delta_X$ is a field-space 1-form, $L_{\delta X}$ increases the degree by one and hence unlike $L_V$ obeys an \emph{anti}-Liebniz rule.
This leads to the identity:
\begin{equation} \label{straightLdeltaX}
L_{\delta_X} = \mathcal{L}_{\delta_X} - I_{\delta(\delta_X)}.
\end{equation}
To see this note that on a scalar $f$ we have (as before)
\begin{equation}
L_{\delta_X} f = I_{\delta_X} \delta f - \delta (I_{\delta_X} f) = \mathcal{L}_{\delta_X} f.
\end{equation}
However since $\mathcal{L}_{\delta_X}$ and $\delta$ are both antiderivations, we have
\begin{equation}
L_{\delta_X} (\delta \phi) = I_{\delta_X} (\delta \delta \phi) - \delta (I_{\delta_X} \delta \phi) = - \delta (\mathcal{L}_{\delta_X} \phi) = \mathcal{L}_{\delta_X} \delta \phi - \mathcal{L}_{\delta (\delta_X)} \phi = (\mathcal{L}_{\delta_X} - I_{\delta (\delta_X)} ) \delta \phi.
\end{equation}
Thus \eqref{straightLdeltaX} holds for field-space $0$-forms and coordinate $1$-forms.
We can now extend this to all forms, using the fact that $L_{\delta_X}$, $\mathcal{L}_{\delta_X}$ and $I_{\delta \delta_X}$ are all antiderivations, and hence satisfy a graded Leibniz rule.
Thus the relation \eqref{straightLdeltaX} is proved for all $(p,q)$ forms.

\section{Variation of the pullback} \label{appendix:pullback}

Let $X: M \to M$ be a diffeomorphism of spacetime, $T$ be a spacetime tensor and field-space $p$-form, and let $X^*T$ denote the pullback under diffeomorphism.
The purpose of this appendix is to prove the following variational formula:
\begin{equation} \label{pullback}
\delta X^*T = X^*(\delta T + \mathcal{L}_{\delta_X} T).
\end{equation}
To prove this formula we first establish it on spacetime tensors, and then on a basis of variational $1$-forms.
We can then extend to arbitrary tensor-valued variational forms by the product rule, using the fact that both $\delta$ and $\mathcal{L}_{\delta_X}$ act as derivations on spacetime tensors, and as antiderivations on variational forms.

To prove \eqref{pullback} for spacetime tensors, it suffices to prove it for scalars, vectors and covectors and then to extend to tensors of arbitrary rank by taking products.
For a scalar function $f$:
\begin{equation}
\delta X^* f = \delta f(X) = (\delta f)(X) + \delta X^A \partial_A f(X) = X^*( \delta f + \delta_X^a \partial_a f) = X^*( \delta f + \mathcal{L}_{\delta_X} f)
\end{equation}
where we have used that $\delta_X^A = \delta X^A \circ X^{-1}$.
Here we temporarily introduce upper-case letter indices on $X^A$ to distinguish them from the lower-case indices on $x^a$.
For a vector $v^a$:
\begin{align}
\delta X^* v^a &= \delta \left( \frac{\partial x^a}{\partial X^A} v^A(X)  \right) \nonumber \\
&= \frac{\partial x^a}{\partial X^A} (\delta v^A)(X) -\frac{\partial x^a}{\partial X^B} \frac{\partial \delta X^B}{\partial x^c} \frac{\partial x^c}{\partial X^A} v^A(X) +  \frac{\partial x^a}{\partial X^A} \delta X^B \partial_B v^A(X) \nonumber \\
&= X^*( \delta v^a - \partial_b \delta_X^a v^b + \delta_X^b \partial_b v^a) \nonumber \\
&= X^*( \delta v^a + \mathcal{L}_{\delta_X} v^a)
\end{align}
For a covector $w_a$:
\begin{align}
\delta X^* w_a &= \delta (\partial_a X^A w_A(X)) \nonumber \\
&= \partial_a X^A (\delta w_A)(X) + \partial_a \delta X^A w_A(X) + \partial_a X^A \delta X^B \partial_B w_A(X) \nonumber \\
&= X^* ( \delta w_a + \partial_a \delta_X^{b} w_b + \delta_X^b \partial_b w_a)\nonumber \\
&= X^* ( \delta w_a + \mathcal{L}_{\delta_X} w_a).
\end{align}
This establishes the identity for all spacetime tensors, using the fact that both $\delta$ and $\mathcal{L}_{\delta_X}$ are derivations.

To extend the identity to variational forms, it suffices to prove it on a basis of variational $1$-forms.
Our basis of variational $1$-forms will consist of $\delta g_{ab}$ and $\delta_X^a$.
For the metric variation $\delta g_{ab}$:
\begin{align}
\delta X^* \delta g_{ab} &= \delta (\partial_a X^A \partial_b X^B \delta g_{AB}(X)) \nonumber \\
&= \partial_a \delta X^A \partial_b X^B \delta g_{AB}(X) + 
\partial_a X^A \partial_b \delta X^B \delta g_{AB}(X) +  
\partial_a X^A \partial_b X^B \delta X^C \partial_C \delta g_{AB}(X) \nonumber \\
&= X^*( (\partial_a \delta_X^c) \delta g_{cb} + (\partial_b \delta_X^c) \delta g_{ac} + \delta_X^c \partial_c \delta g_{ab})
\nonumber \\
&= X^*( \mathcal{L}_{\delta_X} \delta g_{ab}).
\end{align}
And for $\delta_X^a$:
\begin{align}
\delta X^* (\delta_X^a) &= \delta \left( \frac{\partial x^a}{\partial X^A} \delta X^A \right) \nonumber \\
&= - \frac{\partial x^a}{\partial X^B} \frac{\partial \delta X^B}{\partial x^c} \frac{\partial x^c}{\partial X^A} \delta X^A \nonumber \\
&= X^*(\delta_X^b \partial_b \delta_X^a) \nonumber \\
&= X^*(\tfrac12 [\delta_X,\delta_X]^a) \nonumber \\
&= X^*(\delta (\delta_X^a) + \mathcal{L}_{\delta_X} \delta_X^a)
\end{align}
where we have used $\delta (\delta_X) = -\tfrac12 [\delta_X, \delta_X]$ and $\mathcal{L}_{\delta_X} \delta_X = [\delta_X, \delta_X]$.

We can now extend the rule \eqref{pullback} to arbitrary tensor valued forms by induction. 
Suppose that $\alpha$ and $\beta$ each satisfy \eqref{pullback}, their product must also:
\begin{align}
\delta X^*(\alpha \beta) &= \delta (X^*(\alpha) X^*(\beta)) \nonumber \\
&= \delta X^* (\alpha) X^* (\beta) + (-1)^p X^* (\alpha) \delta X^*(\beta) \nonumber \\
&= X^*(\delta \alpha + \mathcal{L}_{\delta_X} \alpha) X^*\beta + (-1)^p X^*(\alpha) X^*(\delta \beta + \mathcal{L}_{\delta_X} \beta) \nonumber \\
&= X^*[(\delta \alpha + \mathcal{L}_{\delta_X} \alpha) \beta + (-1)^p \alpha (\delta \beta + \mathcal{L}_{\delta_X} \beta)] \nonumber \\
&= X^*[(\delta + \mathcal{L}_{\delta_X}) (\alpha \beta)]
\end{align}
This follows from the fact that both $(X^{-1})^* \circ \delta \circ X^*$ and $\delta + \mathcal{L}_{\delta_X}$ are antiderivations, and that an antiderivation is defined by its action on a basis of $1$-forms.
Thus \eqref{pullback} is proved for all tensor-valued variational forms.

\end{document}